\documentclass{aa} %[traditabstract]
\usepackage[varg]{txfonts}
\usepackage{graphicx}
\usepackage{graphics}
\usepackage{amssymb}
\usepackage{amsmath}
\usepackage[T1]{fontenc}
\usepackage[utf8]{inputenc}
\usepackage{pdflscape} %for landscape tables
\usepackage{color} %in order to have colored text
\usepackage[switch]{lineno}
\usepackage{multicol}
\newcommand{\td}[1]{\, \mathrm{d} #1 \,}
\newcommand{\intl}{\int\limits}

\newcommand{\est}[3]{\left( \frac{#1}{#2} \right)^{#3}}
\newcommand{\g}{\ensuremath{\gamma}}
\newcommand{\p}{^{\prime}}

\newcommand{\E}[1]{\times 10^{#1}}
\newcommand{\change}[1]{#1} %so that the ``change'' command must not be deleted everywhere
\begin{document}
\title{The ablation of gas clouds by blazar jets}
\author{Jonathan Heil$^1$, Michael Zacharias$^{1,2}$\thanks{Now at: Laboratoire Univers et Théories, Observatoire de Paris, Université PSL, CNRS, Université de Paris, 92190 Meudon, France}}
\institute{$^1$ Ruhr Astroparticle and Plasma Physics Center (RAPP Center), Insitut f\"ur Theoretische Physik IV, Ruhr-Universit\"at Bochum, \\ D-44780 Bochum, Germany, hejo@tp4.rub.de \\
$^2$ Centre for Space Research, North-West University, Potchefstroom 2520, South Africa, mzacharias.phys@gmail.com}
\date{Received ? / accepted ? }
\abstract{Flaring activity in blazars can last for vastly different time-scales, and may be the result of density enhancements in the jet flow that result from the intrusion of an interstellar cloud into the jet.} %Context, can be empty
{We investigate the lightcurves expected from the ablation of gas clouds by the blazar jet under various cloud and jet configurations.} %Aims
{We derive the semi-analytical formulae describing the ablation process of a hydrostatic cloud, and perform parameter scans of artificial set-ups over both cloud and jet parameter spaces. We then use parameters obtained from measurements of various cloud types to produce lightcurves of these cloud examples.} %Methods
{The parameter scans show that a vast zoo of symmetrical lightcurves can be realized. Both cloud and emission region parameters significantly influence the duration, and strength of the flare. The scale height of the cloud is one of the most important parameters, as it determines the shape of the lightcurve. In turn, important cloud parameters can be deduced from the observed shape of a flare. The example clouds result in significant flares lasting for various time scales.} %Results
{} %Conclusions, can be empty
\titlerunning{Cloud ablation by relativistic jets}
\authorrunning{Heil \& Zacharias}
\maketitle
%
%#######################################################################################################################################
%#######################################################################################################################################
%
\section{Introduction}
Relativistic jets produce a vast range of flaring activity that can last from a few minutes, as in the case of gamma-ray bursts, up to months or years, as in the case of active galactic nuclei. Jets and their flares are most easily studied in blazars, where the jet of the active galaxy points towards Earth \citep{up95} and persists over long time scales. Blazars, as all relativistic jets, exhibit flares that differ substantially in duration and evolution \citep[e.g.,][and references therein]{z18}. A common scenario is a change in the plasma flow across a shock within the jet \citep{mg85}. However, the details of the variation of the plasma parameters that fit the lightcurves, have to be set arbitrarily. Most easily, a variation in the plasma density can account for a flare, and the modulation in the injection results in the specific lightcurve profile. Natural sources of density fluctuations are either a variable particle injection process at the base of the jet, possibly coupled to variations in the accretion disk, or through pick-up of material while the jet moves through the host galaxy. For the latter process, material can be supplied by the interstellar gas, stellar astrospheres, supernova remnants, etc.

\cite{zea17,zea19} used the injection of pick-up material to explain the bright, and long-lasting flare of the blazar CTA~102. In their model, a gas cloud approached the jet and was subsequently ablated. In their picture, the cloud slowly intruded into the jet implying a smoothly varying injection of particles. The number of particles injected into the jet flow depended on the geometry and the density structure of the cloud with few particles being injected in the beginning and at the end of the process and most particles being injected when the center of the cloud intruded the jet.

The successful reproduction of the CTA~102 flare is reassuring. However, in order to understand the full potential of the model, it is necessary to study the influence of various parameters from both cloud and jet on the lightcurve. Therefore in this paper, we revisit the cloud-ablation model from a theoretical point-of-view. We will first explore the requirements of cloud and jet for the ablation to proceed, followed by deriving the time evolution of the particle injection. This is described in Sec.~\ref{sec:ablation} along with a general discussion of the model. We then study the lightcurves of both theoretical (Sec.~\ref{sec:param}) and a few exemplary real (Sec.~\ref{sec:clouds}) clouds. The results are discussed in Sec.~\ref{sec:sumcon}. 

%
%#######################################################################################################################################
%#######################################################################################################################################
%
\section{The cloud ablation model} \label{sec:ablation}
Clouds, like those of the broad-line region (BLR) but also stars and their astrospheres, penetrating the relativistic jet of an AGN have been explored in various applications. Generally, the time-dependent intrusion of the cloud into the jet results in a similar time-dependency of the particle injection into the jet flow. The destruction of the cloud is a consequence of the relativistic jet flow and the associated ram pressure.

The time scale of the ablation process is governed by the speed of the shock that is formed in the cloud as it starts penetrating the jet, i.e. the time the shock needs to cross the cloud. For a rough estimate, one can assume that the shock speed is roughly $c/4$,\footnote{In the downstream frame, the speed of a strong shock is $v_s=u_1/4$. The upstream speed is $u_1\sim c$, i.e. the speed of the jet.} where $c$ is the speed of light. The time for the shock to pass the cloud is $t_s=8R/c$, with the cloud radius $R$. If the intrusion time of the cloud into the jet $t_{in}=2R/v$, where $v$ is the cloud speed, is shorter than $t_s$, the cloud may penetrate deep into the jet before the shock has crossed the object. In this case, the cloud material will be shocked in one instance resulting in a violent burst of particles and radiation \citep{abr10}. This provides the ingredients for fast flares \citep[e.g.,][]{bea12,bba12}.

\begin{figure}
\centering
\includegraphics[width=0.40\textwidth]{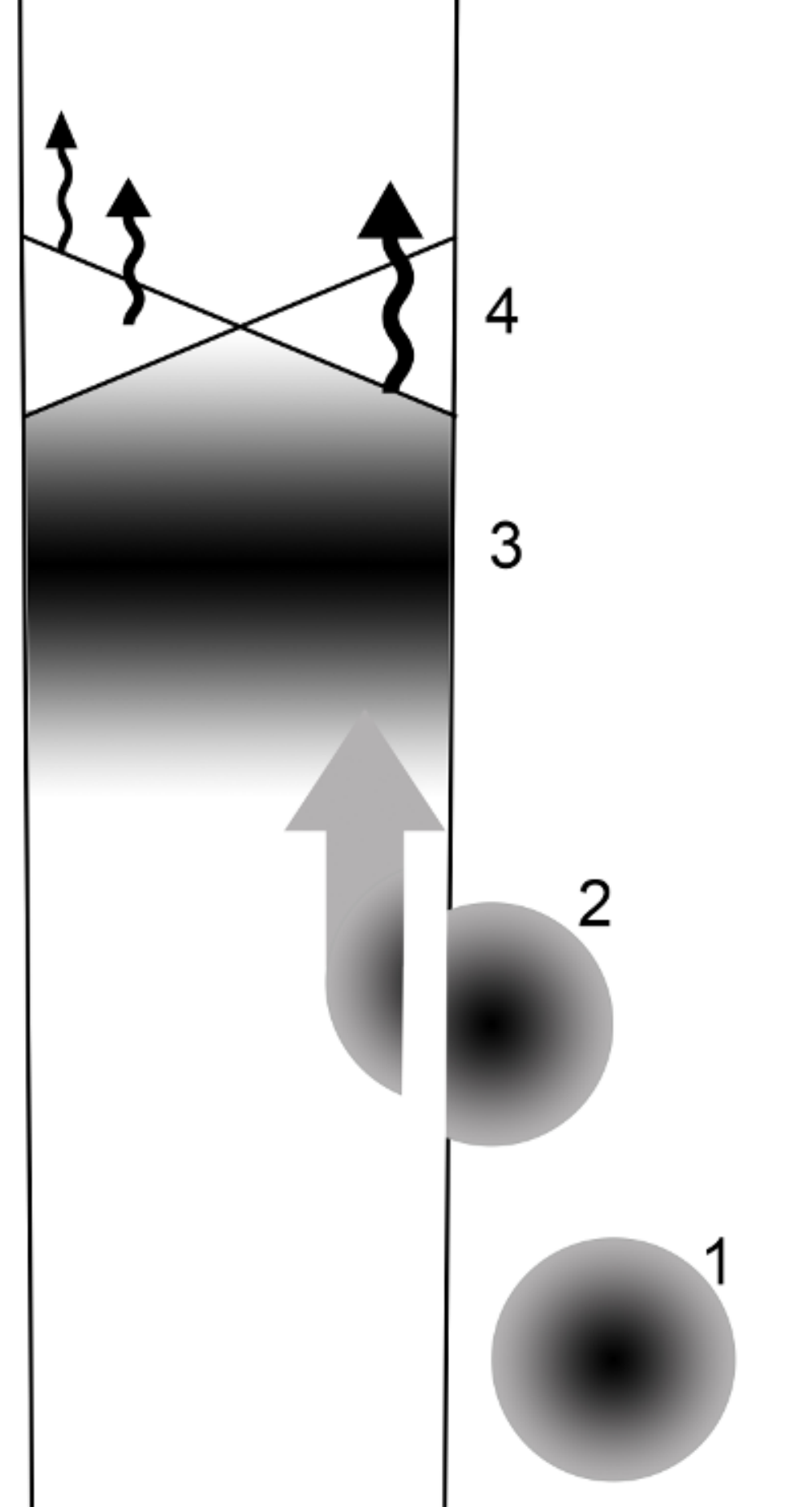}
\caption{Sketch of the ablation process (not to scale). (1) The cloud approaches the jet and (2) is ablated slice-by-slice while entering the jet. (3) The ablated cloud material is mixed into the jet flow resulting in a specific density enhancement. (4) At a downstream shock, the particles are accelerated to non-thermal energies and radiate.
}
\label{fig:model}
\end{figure}
On the other hand, if $t_s<t_{in}$, the shock mostly interacts with the subvolume of the cloud that has already penetrated the jet. In turn, the ablation process is gradual \citep{zea17} as depicted in Fig.~\ref{fig:model}. Simulations by \cite{pbb17} involving a jet-star interaction suggest that the ablation of the astrosphere already begins in the transition layer at the edge of the jet and proceeds while the star continues to penetrate. The ablated material is mixed into the jet flow, where it is accelerated to the jet's bulk speed. The acceleration to non-thermal energies may take place at a downstream shock, such as a recollimation shock \citep{mea08}. Such shocks are ubiquitous in blazar jets \citep{jea17}. Radiative processes \citep{tb19} may occur close to the shock region. At this point we ignore the possibility that shock waves are formed in the jet through the cloud intrusion itself \citep{br12,pbb17}. The potential variations in shock strength and size throughout the jet-cloud interaction cannot be easily quantified, and require in-depth (magneto-)hydrodynamic (MHD)  simulations, which are beyond the scope of this paper.  

Following the gradual ablation and injection of particles into the jet flow, the subsequent acceleration of the particles and emission of radiation proceed similarly. A resulting flare evolves smoothly and is probably symmetric in time,\footnote{Note that this remark concerns the long-term behavior of the flare. On shorter time-scales, substructures in the cloud or instabilities in the jet due to the ablation process, may still trigger shorter spikes in the lightcurve.} 
as the change in the particle injection dominates the lightcurve, while other jet parameters -- such as shock radius -- probably remain fixed. 

In this paper, we focus on the second scenario, where a cloud enters the jet, and is slowly ablated leading to a long-lasting flux enhancement. In order to calculate the amount of ablated material at a given instant of time, the cloud's geometry and density structure are required. While the geometry has a significant influence on the ablated volume, we assume only a spherical geometry for ease of computation. Furthermore, we only consider leptonic radiation processes. While the hadronic components of the cloud are carried along in the jet, we assume that they remain energetically cold and do not participate in the radiation processes, \change{which is a common -- though debated -- assumption in blazar modeling \citep[e.g.,][]{b07,gea11,bea13,zb15,cea15,cea17,Hea19}}. 
Below, we describe the cloud's density structure, the jet condition to ablate the cloud, and the resulting injection function in turn.

%#######################################################################################################################################
\subsection{Cloud density}
An isothermal cloud is bound by its own gravitational pull. If we ignore any external forces that may shape it into different structures, the cloud is spherically symmetric with radial coordinate $r$. As discussed in \cite{zea17}, the cloud's density structure $n(r)$ is defined by the equation of hydrostatic equilibrium

\begin{align}
    k_B T \frac{\td{n(r)}}{\td{r}} = -4\pi \frac{Gm_p^2 n(r)}{r^2} \intl_0^{r} \td{\bar{r}} \bar{r}^2 n(\bar{r}) \label{eq:hydro}
\end{align}
with the temperature $T$, Boltzmann's constant $k_B$, proton mass $m_p$ and gravitational constant $G$. We have assumed that the cloud predominantly consists of hydrogen. 

With some manipulations of Eq.~(\ref{eq:hydro}), one reaches the Lane-Emden equation, which does not provide an analytical solution in this case. However, from the asymptotic solution $n\propto r^{-2}$, as well as the boundary conditions $n(0)<\infty$ and $\left.\td{n}/\td{r}\right|_{r=0}=0$, one can derive a reasonable approximation \citep{zea17,bbea18}:

\begin{align}
 n(r) = \frac{n_0}{1+(r/r_0)^2} \label{eq:clouddens},
\end{align}
with the central density $n_0$, and the scale height 

\begin{align}
 r_0 &= \sqrt{\frac{2k_B T}{4\pi G m_p^2 n_0}} = \sqrt{\tilde{c}\frac{T}{n_0}} \nonumber \\
 &= 4\E{12} \est{T}{140\,\mbox{K}}{1/2} \est{n_0}{10^{15}\,\mbox{cm}^{-3}}{-1/2}\,\mbox{cm} \label{eq:scaleheight},
\end{align}
with $\tilde{c} := 2k_B/(4\pi G m_p^2)=1.17\E{38}\,$cm$^{-1}$K$^{-1}$. Note that the scale height in \cite{zea17} contains a minor calculation error on the order of unity, which has been corrected here.
As clouds cannot be infinitely large, we define an outer radius $R$ after which the density is set to zero.

As the isothermal cloud contains predominantly hydrogen, the sound speed simply is $c_s=(5k_BT/3m_p)^{1/2}$. Using Eq.~(\ref{eq:scaleheight}), this becomes

\begin{align}
    c_s &= \sqrt{\frac{5k_B n_0r_0^2}{3m_p\tilde{c}}} \nonumber \\
    &= 1.4\E{5} \est{n_0}{10^{15}\,\mbox{cm}^{-3}}{1/2} \est{r_0}{4\E{12}\,\mbox{cm}}{}\, \mbox{cm}\,\mbox{s}^{-1} \label{eq:soundspeed},
\end{align}
which shows that all relevant speeds are much larger than the cloud's sound speed. This verifies \textit{a posteriori} that large Mach numbers are achieved and a strong shock is formed in the cloud. 

%
%#######################################################################################################################################
\subsection{Necessary jet condition} 
The ablation process commences, if the cloud's gravitational pull cannot withstand the jet's ram pressure. While the details of the process require (M)HD simulations, which are beyond the scope of this paper, we can provide a rough estimate on the necessary jet condition to ablate the cloud.

In the frame of the host galaxy, the relativistic jet containing a fraction of $a$ cold protons and $(1-a)$ positrons per electron, exerts the ram pressure

\begin{align}
    P_R \approx \Gamma (\Gamma-1) a m_p c^2 n_j \label{eq:rampressure},
\end{align}
with the bulk Lorentz factor $\Gamma$, the proton rest energy $m_pc^2$, and the jet's electron density $n_j$. We assumed that $am_p>\bar{\gamma}m_e$, where $\bar{\gamma}$ is the average electron Lorentz factor. This inequality implies that the mass of protons is greater than the average relativistic mass of the electrons. In turn, the ram pressure is dominated by the protons. This approximations is improved, if the protons have non-negligible kinetic energy in the comoving frame.

The ram pressure must overcome the cloud's pressure on its particles. Following the hydrostatic equilibrium condition, the cloud's pressure is

\begin{align}
    P_c(r) = \frac{n_0 k_B T}{1+\left( r/r_0 \right)^2} \approx n_0k_B T \label{eq:gravpressureI},
\end{align}
where the approximation holds within the cloud's center ($r\ll r_0$).

If the ram pressure is larger than the cloud's central pressure, the cloud will be destructed entirely. Setting $P_R>P_c(r\ll r_0)$, and solving for the bulk Lorentz factor results in

\begin{align}
    \Gamma(\Gamma-1) &> \frac{n_0 k_B T}{am_p c^2 n_j} \nonumber \\
    &= 128 \est{n_0}{10^{15}\,\mbox{cm}^{-3}}{} \est{T}{140\,\mbox{K}}{}  \est{a}{0.1}{-1} \est{n_j}{10^3\,\mbox{cm}^{-3}}{-1} \label{eq:requiredbulk} .
\end{align}
Taking the square root provides us with the required value for the bulk Lorentz factor of $\Gamma\gtrsim 12$, which is achieved in many blazar jets \citep[e.g.,][]{jea17}. Hence, a cloud with the provided parameters is indeed destructed. Parts of a denser cloud, or one with a higher temperature, may possibly survive the encounter. Note that the chances of the cloud's survival are much reduced, if the jet protons are not cold. As a side note: while stars may be stripped of their astrospheres, the star itself should survive the encounter with a relativistic jet.

Additionally, we can consider the time the cloud needs to cross the jet. As the central region of the cloud within the scale height is the densest part of the cloud, we consider this size in the following estimate. The ablation process is governed by the shock that forms during the interaction. Hence, the crossing time of the shock through the cloud is a good estimator of the ablation time, as it disrupts the internal structure of the cloud, adding (or enhancing) the turbulent motions in the cloud, which weakens the gravitational pull. We can derive the minimum shock speed $v_s$ required to cross the cloud, before the cloud with speed $v$ has crossed the jet of radius $R_j$:

\begin{align}
    t_{\rm cross} = \frac{2R_j}{v} &< t_{s} = \frac{2r_0}{v_s} \nonumber \\
    \Leftrightarrow v_s &< \frac{r_0}{R_j} v \nonumber \\
    &= 3.2\E{3} \est{r_0}{4\E{12}\,\mbox{cm}}{} \est{v}{2\E{7}\,\mbox{cm\,s}^{-1}}{} \nonumber \\
    &\quad\times \est{R_j}{2.5\E{16}\,\mbox{cm}}{-1}\,\mbox{cm\,s}^{-1}. \label{eq:shockspeed1}
\end{align}
If we express the jet radius as a function of distance $z_j$ from the black hole using $R_j = z_j\tan{\Gamma^{-1}}\approx z_j/\Gamma$, Eq.~(\ref{eq:shockspeed1}) becomes

\begin{align}
    v_s &< \frac{r_0\Gamma}{z_j}v \nonumber \\
    &= 1.3\E{3} \est{r_0}{4\E{12}\,\mbox{cm}}{} \est{v}{2\E{7}\,\mbox{cm\,s}^{-1}}{} \nonumber \\
    &\quad\times \est{\Gamma}{10}{} \est{z_j}{6.5\E{17}\,\mbox{cm}}{-1}\,\mbox{cm\,s}^{-1}. \label{eq:shockspeed2}
\end{align}
Comparing this to our earlier estimate that the shock speed may actually be on the order of $~\sim c/4$, the cloud will not be able to cross the jet in time. This changes for clouds close to the base of the jet, where the speed of motion of the cloud is a lot higher, and the jet a lot narrower.

The gradual ablation of the cloud results in the injection of the cloud particles into the jet, where they get mixed into the bulk flow. The injection function is derived in the next section.

%#######################################################################################################################################
\subsection{Injection function into the jet emission region}
The number of particles $\td{N}$ entering the jet in a given time step $\td{t}$, depends on the density of the cloud and the ablated volume $\td{V}$ of a slice of the cloud. As in \cite{zea17}, we denote with $x=0$ the point of the cloud that first touches the jet, with $x=R$ the centre of the cloud, and with $x=2R$ the far side of the cloud. The ablated volume at position $x$ becomes \citep{zs13}

\begin{align}
 \td{V}(x) = \td{x}\int^{x} \td{A}(\tilde{x}) = \pi (2Rx-x^2)\td{x} \label{eq:ablvol},
\end{align}
with the width $\td{x}$ and the cross-section $A(x)$ of the ablated volume. The particle number in a slice yields

\begin{align}
    \td{N}(x) &= \td{x}\int^{x} n(r) \td{A}(\tilde{x}) \nonumber \\
    &= \pi n_0 \td{x}  r_0^2 \ln{\left( \frac{r_0^2 + R^2}{r_0^2 + (R-x)^2} \right)}       \label{eq:particleabl}
\end{align}

If the cloud enters the jet with constant speed $v$, the length scales can be transformed to time scales. Then the number of particles entering the jet in a given time step $\td{t}=\td{x}/v$ is

\begin{align}
    \frac{\td{N}(t)}{\td{t}} = \pi n_0 v r_0^2 \ln{\left( \frac{t_0^2 + t_R^2}{t_0^2 + (t_R-t)^2} \right)}         \label{eq:particlerate} ,
\end{align}
with $t_0 = r_0/v$, and $t_R=R/v$.

In the simulations below, the radiation is calculated in the comoving frame of the jet. Hence, the particle rate, Eq.~(\ref{eq:particlerate}), must be transformed to the comoving frame of the jet.\footnote{Quantities in the comoving frame are denoted with primes.} The jet flows with bulk Lorentz factor $\Gamma$, and we assume that the cloud enters the jet in a right angle in the galactic frame. The Lorentz transformation of the time step is $\td{t}=\Gamma\td{t\p}$, while the transformation of the time coordinate is $t=t\p/\Gamma$ due to the right angle. Then, the particle rate becomes in the comoving frame:

\begin{align}
    \frac{\td{N}(t\p)}{\td{t\p}} = \Gamma\pi n_0 v r_0^2 \ln{\left( \frac{(\Gamma t_0)^2 + (\Gamma t_R)^2}{(\Gamma t_0)^2 + (\Gamma t_R-t\p)^2} \right)}         \label{eq:pratecomov} .
\end{align}
These initially thermal particles get accelerated in the jet through a process which we do not specify here, and are subsequently injected into the emission region (see for more in-depth models, e.g., \citealt{cea12,ws15,bb19}). We assume that a fraction $\epsilon_c\sim 0.1$ \citep[e.g.,][]{ssa13} of the cloud electrons is accelerated and the resulting spectrum is a power-law with index $s\p$ between a minimum and maximum Lorentz factor, $\gamma_{\rm min}\p$ and $\gamma_{\rm max}\p$, respectively. Note again that we assume that the hadronic cloud particles remain energetically cold. The injection luminosity of cloud electrons into the emission region of the jet becomes

\begin{align}
    L\p_{\rm inj,c}(t) &= m_e c^2 \epsilon_c \frac{\td{N}(t\p)}{\td{t\p}} \intl_{\gamma_{\rm min}\p}^{\gamma_{\rm max}\p} \gamma^{\prime 1-s\p}\td{\gamma\p} \nonumber \\
    &= m_e c^2 \epsilon_c \frac{\td{N}(t\p)}{\td{t\p}} \begin{cases}
        \frac{\ln{\left( \gamma_{\rm max}\p/\gamma_{\rm min}\p \right)}}{\gamma_{\rm min}^{\prime -1} - \gamma_{\rm max}^{\prime -1}} & s\p=2 \\
        \frac{1}{2-s\p}\left( \gamma_{\rm max}^{\prime 2-s\p} - \gamma_{\rm min}^{\prime 2-s\p} \right) &\mbox{else}
    \end{cases} \label{eq:injlum}.
\end{align}

While the time dependency in Eq.~(\ref{eq:injlum}) is obviously the same as in \cite{zea17}, here we have also derived the full transformation and the correct normalization factor. These were only indirectly considered or treated as free parameters in \cite{zea17}. Therefore, Eq.~(\ref{eq:injlum}) provides -- within the given assumptions -- the correct particle injection function of a slowly ablated cloud into the emission region of a jet.

%
%#######################################################################################################################################
%#######################################################################################################################################
%
\section{Parameter study} \label{sec:param}
\begin{table*}
\caption{Jet emission region parameter definition, symbol and value for the FSRQ and BL Lac object cases. 
}
\begin{tabular}{lccc}
Definition				        & Symbol 	    		& FSRQ      & BL Lac object \\
\hline
Distance to black hole	        & $z$			    	& $6.5\times 10^{17}\,$cm    & $1.0\times 10^{19}\,$cm \\ 
Doppler factor      	        & $\delta$			    & $35\,$    & $35\,$ \\ 
Emission region radius			& $R_j\p$			    & $2.5\times 10^{16}\,$cm    & $1.0\times 10^{17}\,$cm \\ 
Magnetic field strength	        & $B_j\p$			    & $3.7\,$G  & $1.0\,$G \\ 
e$^{-}$ injection luminosity	& $L_{\rm inj}\p$	    & $2.2\times 10^{43}\,$erg/s & $5.0\times 10^{42}\,$erg/s \\ 
Min. e$^{-}$ Lorentz factor	    & $\gamma_{\rm min}\p$	& $1.3\times 10^1\,$    & $1.6\times 10^2\,$ \\ 
Max. e$^{-}$ Lorentz factor	    & $\gamma_{\rm max}\p$	& $3.0\times 10^3\,$    & $3.0\times 10^6\,$ \\ 
e$^{-}$ spectral index			& $s\p$				    & $2.4\,$   & $2.2\,$ \\ 
Escape time scaling			    & $\eta_{\rm esc}\p$	& $10.0\,$  & $10.0\,$ \\ 
%Acc. to escape time ratio	    & $\eta_{\rm acc}$		& $1.0\,$   & $1.0\,$\\ 
%Accretion Eddington ratio		& $\eta_{\rm Edd}$		& $0.35\,$  & $0.35\,$\\ 
BLR Temperature     	        & $T_{\rm BLR}$ 		& $5.0\times 10^4\,$K   & -- \\ 
Cosmological redshift           & $z_{\rm red}$         & $1.037$   & $1.037$ \\ 
\end{tabular}
\label{tab:jetparam}
\end{table*}
\begin{table}
\caption{Baseline cloud parameter definition, symbol and value for the artificial clouds.
}
\begin{tabular}{lcc}
Definition				 & Symbol 	     & Value \\
\hline
Cloud radius		     & $R$	         & $6.0\times 10^{13}\,$cm \\ 
Cloud scale height	     & $r_0$	     & $4.0\times 10^{12}\,$cm \\ 
Cloud density   	     & $n_0$	     & $1.0\times 10^{15}\,$cm$^{-3}$ \\
Cloud speed 		     & $v$	         & $2.0\times 10^{7}\,$cm/s \\
Acceleration efficiency  & $\epsilon_c$  & $0.1$ \\ 
\end{tabular}
\label{tab:baseparam}
\end{table}
\begin{figure*}
\begin{minipage}{0.49\linewidth}
\centering \resizebox{\hsize}{!}
{\includegraphics{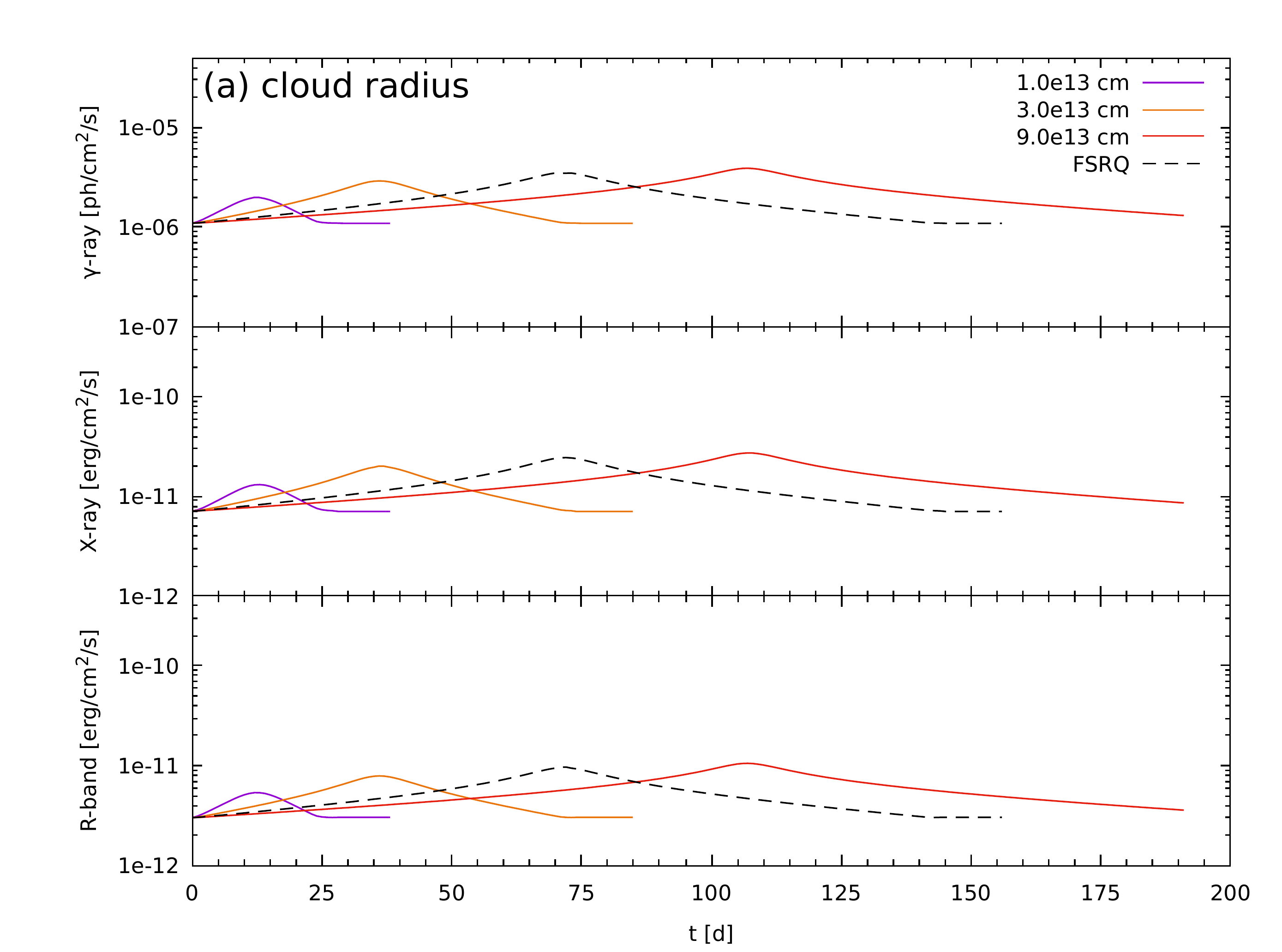}}
\end{minipage}
\hspace{\fill}
\begin{minipage}{0.49\linewidth}
\centering \resizebox{\hsize}{!}
{\includegraphics{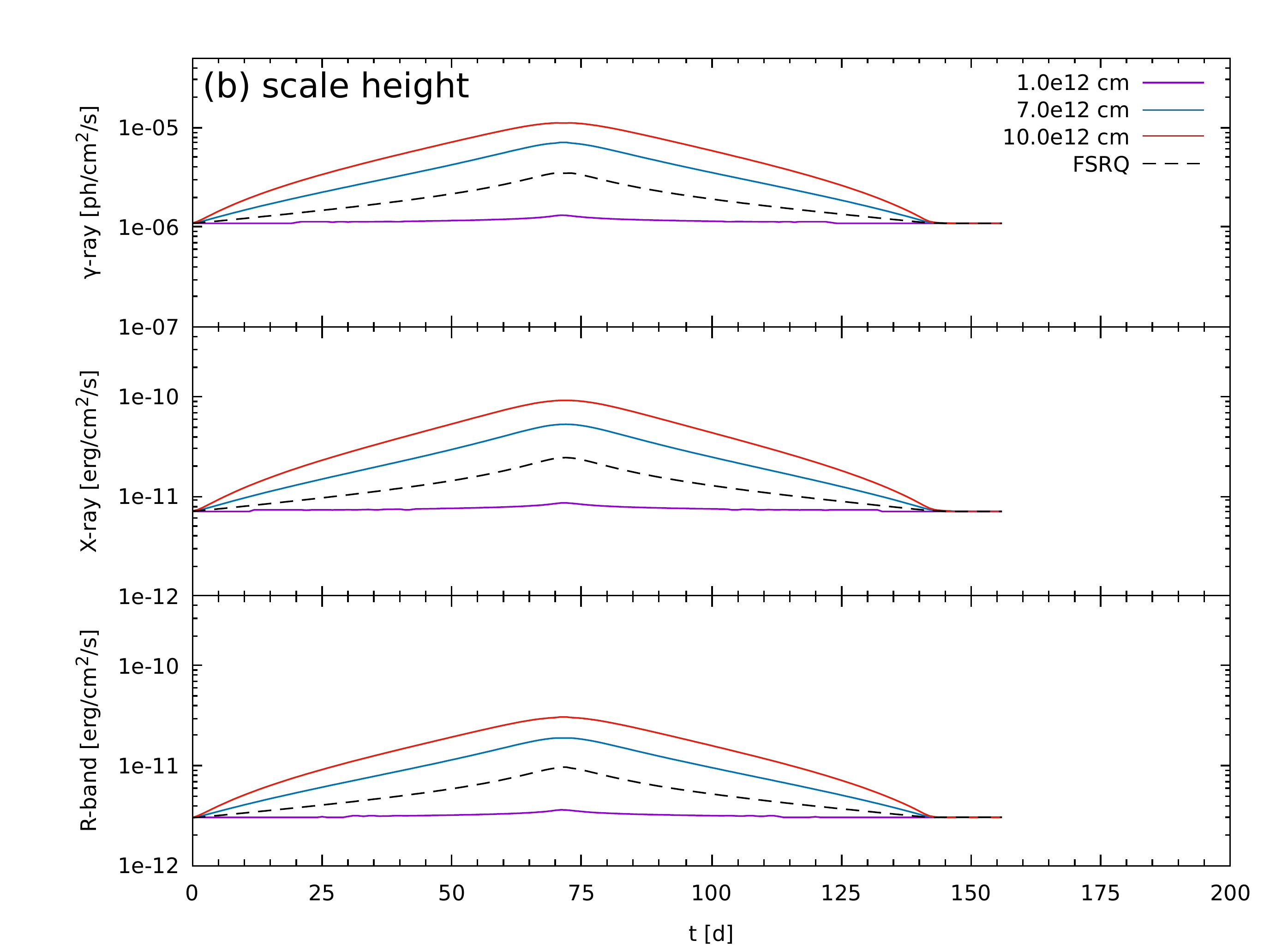}}
\end{minipage}
\newline
\begin{minipage}{0.49\linewidth}
\centering \resizebox{\hsize}{!}
{\includegraphics{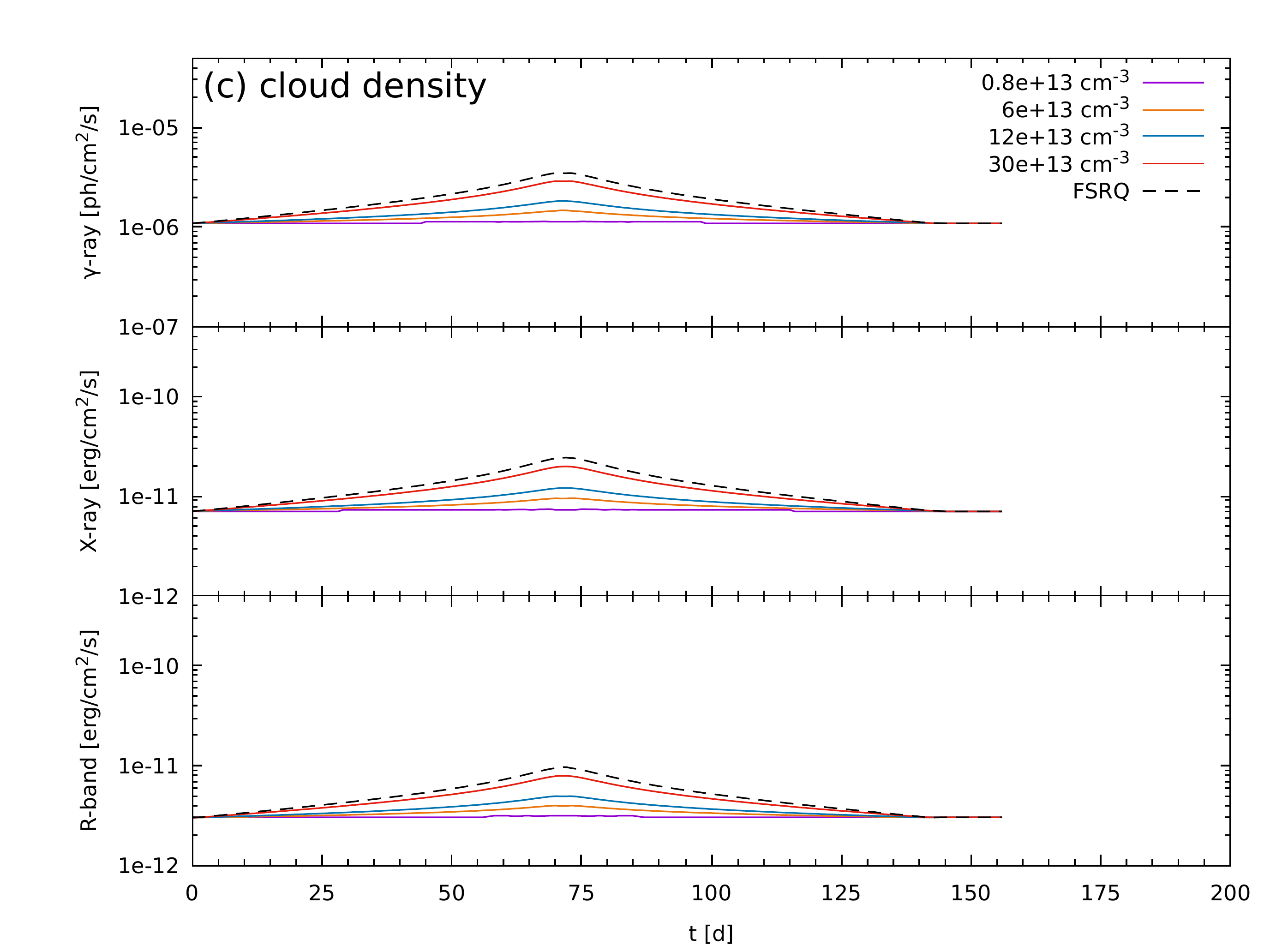}}
\end{minipage}
\hspace{\fill}
\begin{minipage}{0.49\linewidth}
\centering \resizebox{\hsize}{!}
{\includegraphics{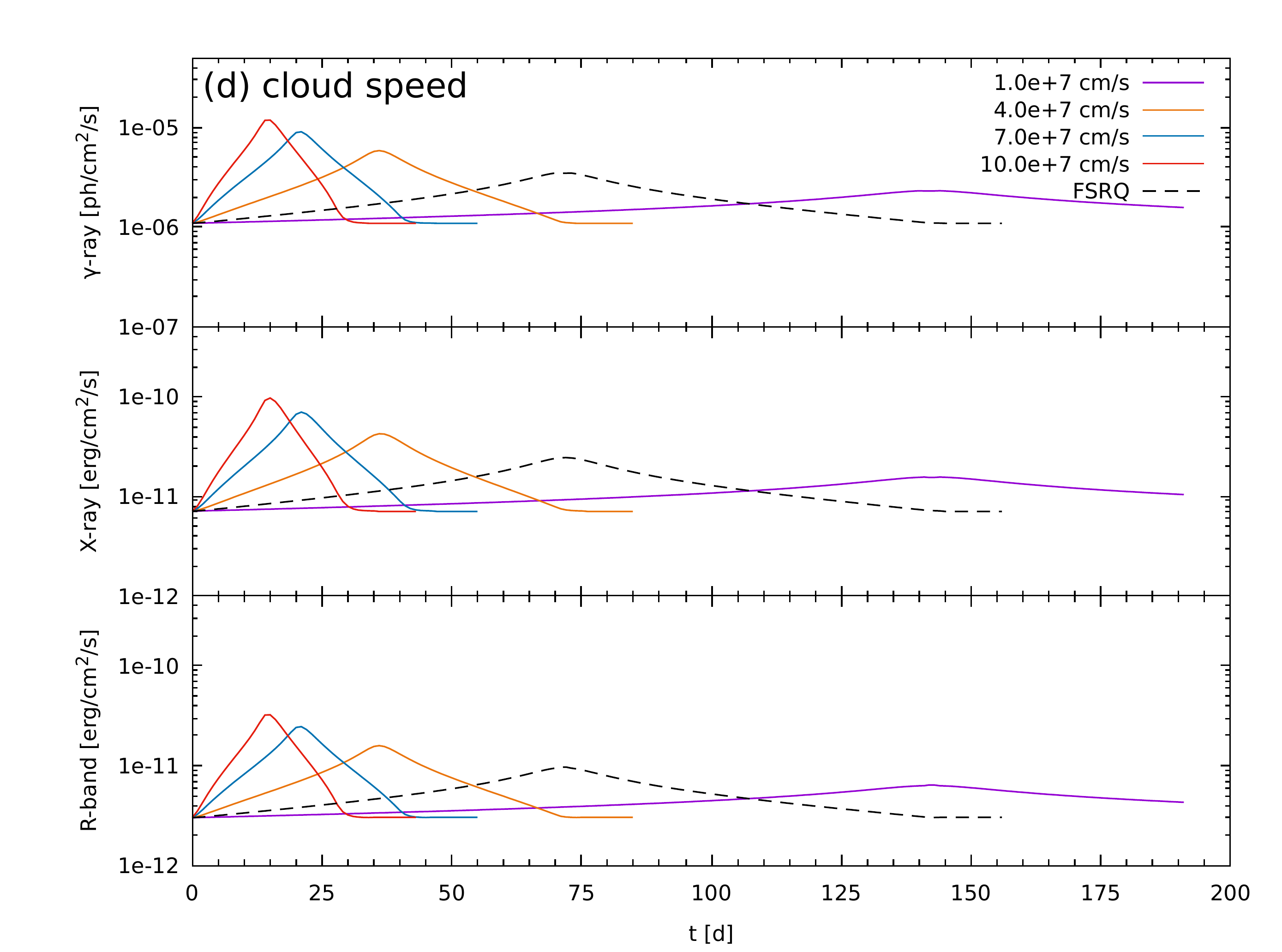}}
\end{minipage}
\caption{Lightcurves in the observer's frame for different parameter values of theoretical clouds. In each panel, lightcurves in the \g-ray, X-ray and R band are shown for different varied parameters: (a) cloud radius $R$, (b) scale height $r_0$, (c) cloud density $n_0$, and (d) cloud speed $v$. The dashed black lightcurve employs the baseline parameters given in Tab.~\ref{tab:baseparam}. Note the logarithmic y-axes.
}
\label{fig:parastud1}
\end{figure*} 
\begin{figure*}
\begin{minipage}{0.49\linewidth}
\centering \resizebox{\hsize}{!}
{\includegraphics{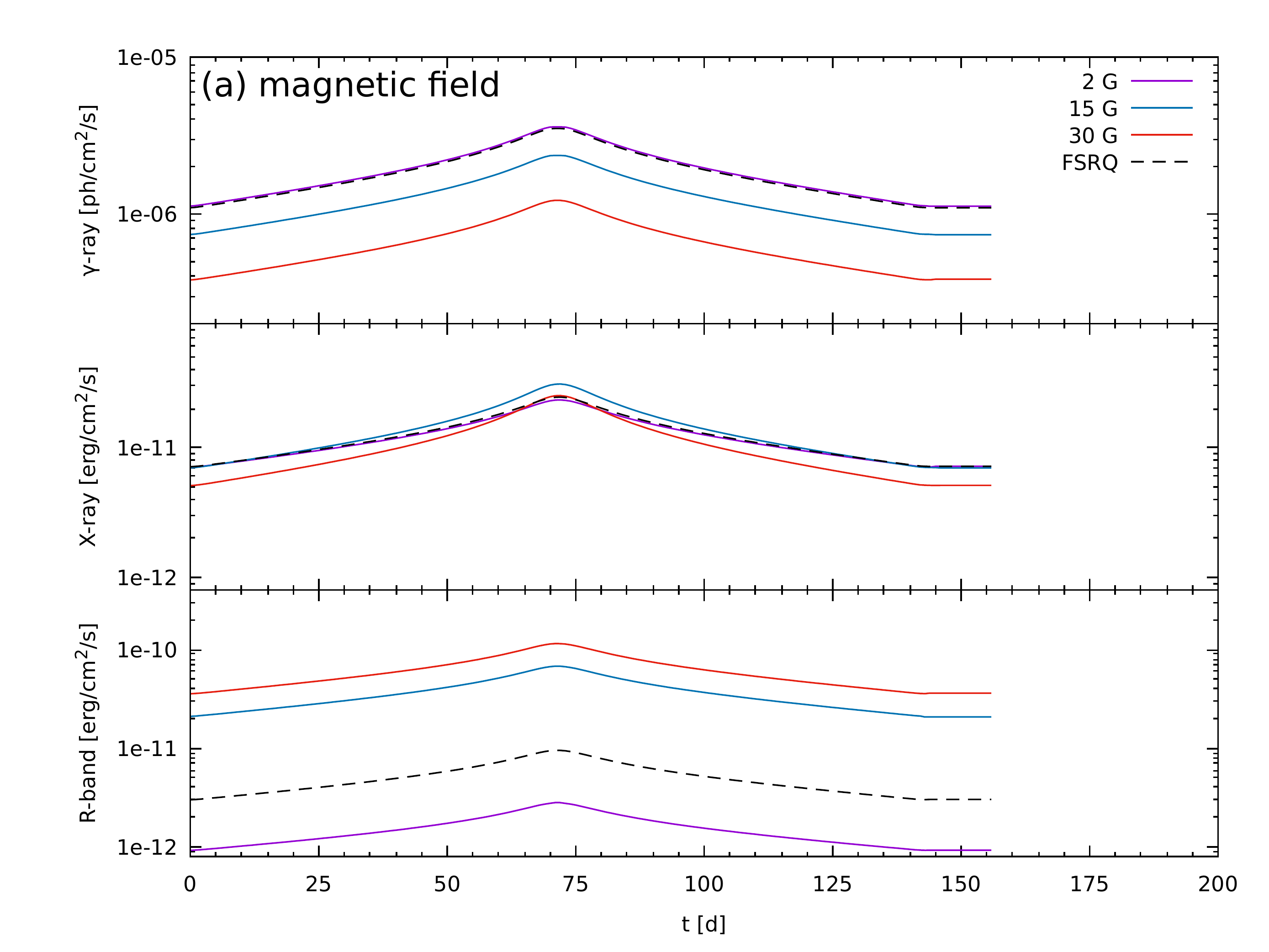}}
\end{minipage}
\hspace{\fill}
\begin{minipage}{0.49\linewidth}
\centering \resizebox{\hsize}{!}
{\includegraphics{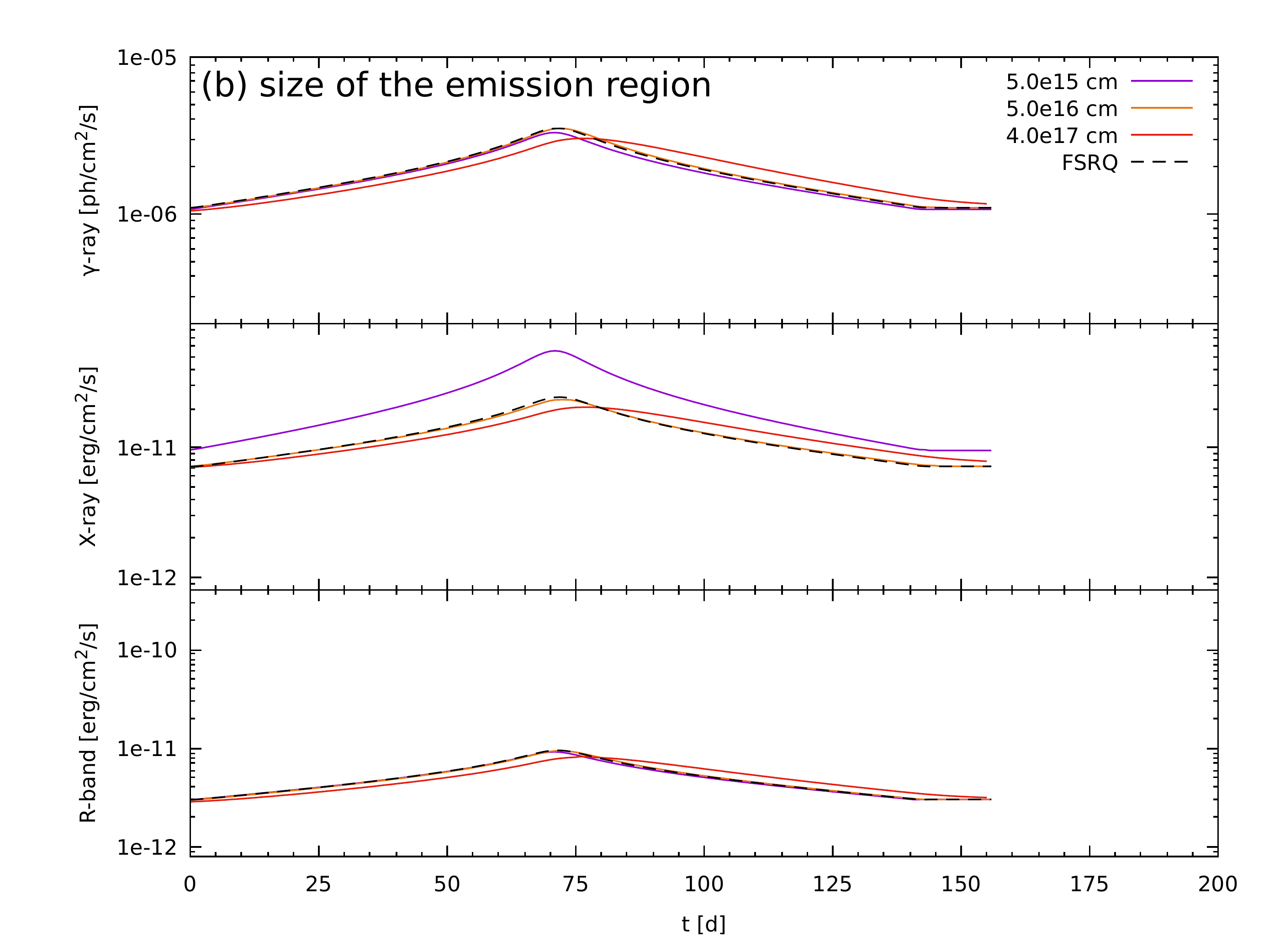}}
\end{minipage}
\newline
\begin{minipage}{0.49\linewidth}
\centering \resizebox{\hsize}{!}
{\includegraphics{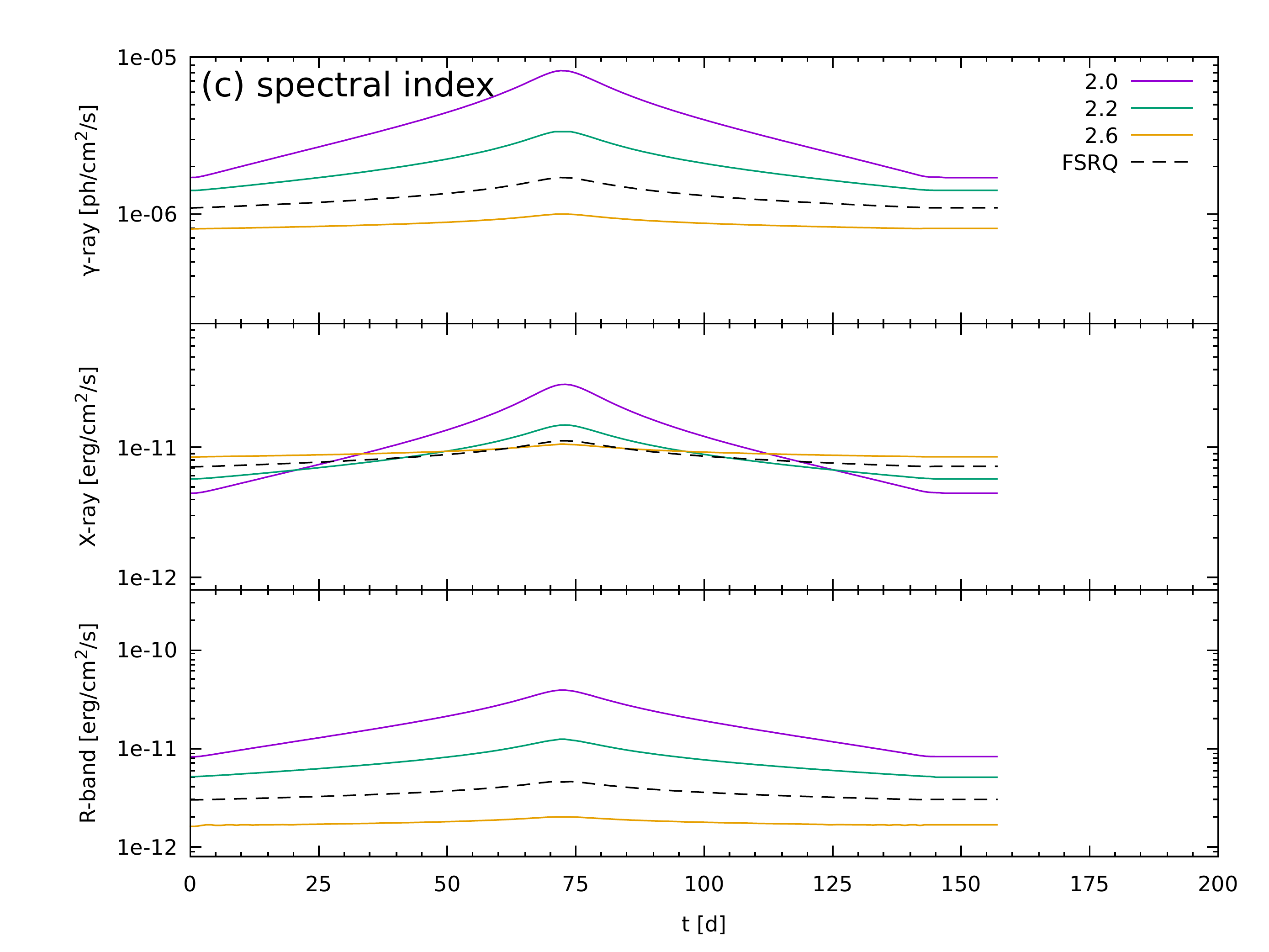}}
\end{minipage}
\hspace{\fill}
\begin{minipage}{0.49\linewidth}
\centering \resizebox{\hsize}{!}
{\includegraphics{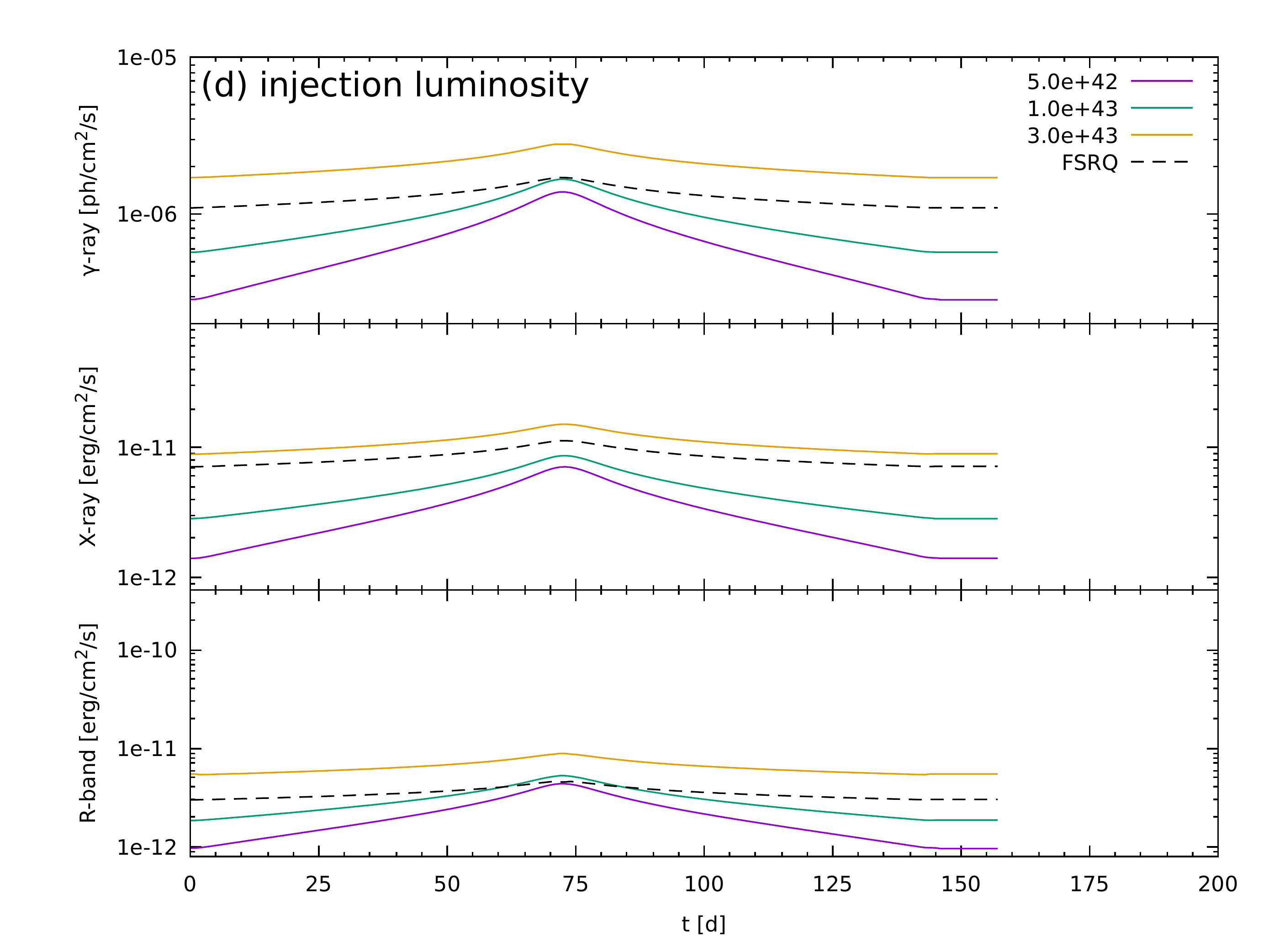}}
\end{minipage}
\caption{Lightcurves in the observer's frame of theoretical clouds for different parameter values of the jet emission region. In each panel, lightcurves in the \g-ray, X-ray and R band are shown for different varied parameters: (a) magnetic field $B\p_j$, and (b) size $R\p_j$ of the emission region, (c) spectral index $s\p$ of the electron distribution, and (d) the injection luminosity $L\p_{\rm inj}$ of the quiescent state. The dashed black lightcurve employs the FSRQ parameters given in Tab.~\ref{tab:jetparam}. Note the logarithmic y-axes.
}
\label{fig:parastudEmReg}
\end{figure*} 
\begin{figure}
\centering
\includegraphics[width=0.48\textwidth]{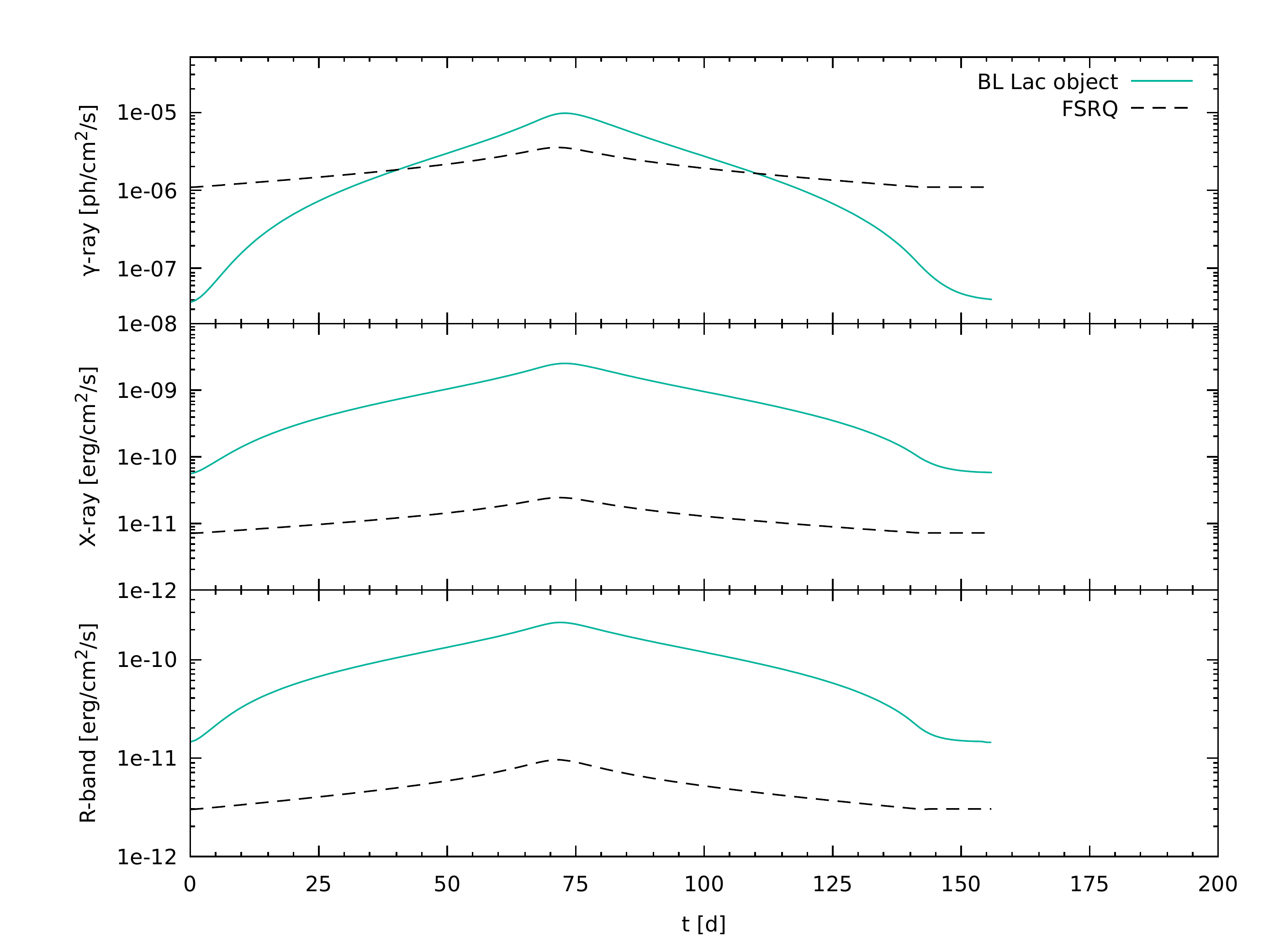}
\caption{Lightcurve in the observer's frame of a theoretical cloud employing the BL Lac object parameter set (turquoise) of the emission region compared to the FSRQ parameter set (dashed black). The parameter sets are given in Tab.~\ref{tab:jetparam}. Note the logarithmic y-axes.
}
\label{fig:parastudSSC}
\end{figure}
In this section we provide a thorough study on the resulting lightcurves following a scan over parameters of the cloud and the emission region. Here and in the following, we use the time-dependent, leptonic one-zone code extensively described in \cite{db14} and \cite{zea17}. 
The emission region parameters are given and defined in Tab.~\ref{tab:jetparam}, providing cases for both flat spectrum radio quasars (FSRQs) and BL Lac objects. To study FSRQs, we use the same parameters as \cite{zea17}. Included emission processes are synchrotron, synchrotron-self Compton (SSC) and inverse-Compton scattering of photons from the accretion disc and the BLR. The parameters provided in Tab.~\ref{tab:jetparam} imply an initial density of non-thermal electrons of $8\E{3}\,$cm$^{-3}$ for FSRQs and $7\E{2}\,$cm$^{-3}$ for BL Lac objects, respectively.\footnote{\change{Note that the seemingly high density in the FSRQ case follows from the modeling of CTA~102 by \cite{zea17}, while the BL Lac case is an adaptation from the FSRQ values. Both values are within bounds found in other studies \citep[e.g.,][]{bea13,zw16}.}} Note that the redshift in Tab.~\ref{tab:jetparam} is required for the flux normalization. 

The baseline cloud parameters, from which we conduct the parameter study of this section, are given in Tab.~\ref{tab:baseparam}. Note that the injection luminosity provided by Eq.~(\ref{eq:injlum}) is added on top of the jet values listed in Tab.~\ref{tab:jetparam}. We assume that the electrons supplied from the ablation process are accelerated to the same spectral behavior -- namely the same minimum and maximum Lorentz factors and spectral index -- as the already existing jet electrons. This is reasonable, if they are accelerated in the same environment at a shock downstream of the jet-cloud interaction site.

For the parameter study each cloud parameter is varied individually resulting in the four plots shown in Fig.~\ref{fig:parastud1}, which display the lightcurves in the \g-ray-, X-ray- and R-band utilizing the FSRQ jet parameters. The results are as follows.

\begin{enumerate}
    \item[(a)] {\it Variation of cloud radius $R$:} As expected, the flare takes longer to evolve with larger $R$, while also the peak flux increases mildly. A larger radius implies overall a larger number of cloud particles explaining the mild increase in flux.
    \item[(b)] {\it Variation of scale height $r_0$:} As $r_0$ governs the size of the region with constant, maximal density $n_0$, the total number of particles significantly changes with a variation of $r_0$. For small $r_0$, the number of particles in the cloud becomes so low that a variation in flux is barely visible (magenta curve). On the other hand, a larger $r_0$ not just increases the flux, but also changes the curvature of the lightcurve owing to the combination of higher particle numbers in a larger cloud volume. In turn, the peak becomes less pronounced with increasing scale height.
    \item[(c)] {\it Variation of density $n_0$:} In this case, the peak fluxes are linearly altered, as the synchrotron and external-Compton\footnote{SSC is negligible in this parameter set.} processes linearly depend on the particle density.
    \item[(d)] {\it Variation of speed $v$:} The influence of $v$ on the lightcurves is involved, as it changes both the normalization factor and the duration of the event. Therefore, slower speeds result in longer, but less pronounced flares in line with the discussion in Sec.~\ref{sec:ablation}.
\end{enumerate}

Obviously, the cloud parameters have a strong influence on the lightcurves, resulting in a zoo of potential solutions, which could explain many symmetrical flares. 

However, the parameters of the emission region itself may also influence the lightcurve. We have tested this by varying individually the magnetic field $B\p_j$, and the size $R\p_j$ of the emission region, as well as the spectral index $s\p$ of the electron distribution and the jet injection luminosity $L\p_{\rm inj}$. The baseline cloud parameters are unchanged. The results are shown in Fig.~\ref{fig:parastudEmReg}, and the details are as follows.

\begin{enumerate}
    \item[(a)] {\it Variation of the magnetic field $B_j\p$:} Obviously, the synchrotron component (optical band) reacts directly to changes in the magnetic field, while the $\gamma$-ray component (external Compton on BLR in this case) remains at the same flux level for most cases and starts to decrease for high magnetic field strengths. This decrease is expected as the synchrotron cooling begins to dominate the external-Compton cooling resulting in a decreased efficiency of the external-Compton process. Similar statements can be made for the X-ray domain exhibiting similar fluxes for most magnetic field values and only deviating for the highest magnetic field strengths. Here, the SSC process starts to dominate the external-Compton process in this energy range.
    \item[(b)] {\it Variation of the size $R_j\p$:} For most values, there is no noteworthy change in the lightcurves. At the smallest size, the SSC process dominates in the X-ray domain due to the increased densities in the synchrotron photons. At the largest size, the dynamical and escape time scales become so long (compared to the chosen time step of $1\,$d in the observer's frame) that particles remain much longer in the emission region, and the fluxes decrease slower than they rise.
    \item[(c)] {\it Variation of the electron spectral index $s\p$:} Following Eq.~(\ref{eq:injlum}), the shape of the accelerated (i.e. injected) particle distribution has a strong influence on the injection luminosity of the cloud particles. While the total number of particles does not change, their influence is distributed to higher energies for harder electron distributions. In turn, the variation is more pronounced. The opposite is true for softer electron distributions.
    \item[(d)] {\it Variation of the injection luminosity $L_{\rm inj}\p$:} This is the injection luminosity of the initial emission region, and its value plays a significant role for the observed variability. For larger values, the luminosity added by the cloud is relatively smaller and the variation does not emerge strongly from the quiescence state. On the other hand, for a small injection luminosity, the cloud injection becomes significant displaying more pronounced variability.
\end{enumerate}

As a final test, we have considered a typical BL Lac object parameter set \citep[e.g.,][]{bea13} with parameters provided in Tab.~\ref{tab:jetparam}. The lightcurve is shown in Fig.~\ref{fig:parastudSSC}. The changes in the parameters of the emission region compared to the FSRQ case concern a larger distance from the black hole in order to avoid inverse-Compton scattering on external photon fields, which are located at smaller distances \citep{zea19}, a larger emission region, a smaller magnetic field, a smaller injection luminosity, and a harder and more energetic electron distribution. Following the discussion on individual changes of emission region parameters, we can expect a significant change in the lightcurve behavior.  

Indeed, the lightcurves in the BL Lac object case differ considerably from the FSRQ case shown as the dashed black line in Fig.~\ref{fig:parastudSSC}, despite using the same cloud parameters. The variation in all energy bands exceeds one order of magnitude, and even two orders of magnitude in the \g-ray band. The latter can easily be understood, as the SSC process depends quadratically on the particle distribution. Hence, as the only change in the cloud injection is the particle number, the SSC flux must change quadratically compared to the synchrotron component. The stronger reaction in the X-ray domain is amplified compared to the FSRQ case, as this energy band is now produced by highly energetic electrons emitting synchrotron emission, which is much more variable than in the baseline case where the X-ray band is dominated by low-energetic electrons producing inverse-Compton emission.

%
%#######################################################################################################################################
%#######################################################################################################################################
%
\section{Interstellar objects} \label{sec:clouds}
\begin{table*}
\caption{Parameters of interstellar clouds. The scale height $r_0$ is calculated from the other parameters and not a free parameter. %The astrosphere is a special case with parameter definitions provided in App.~\ref{app:RGB}.
}
\begin{tabular}{cl|cccc|c|c} %l
& Type			    	& $R$  & $T$   & $n_0$   & $v$ & $r_0$ & Ablate? \\ % & Reference \\
&                       & [cm] & [K]   & [cm$^{-3}$] & [cm\,s$^{-1}$]  & [cm] &  \\ % & \\
\hline
(a) & Giant molecular clouds & $7.7\E{19}$ & $15$ & $2.0\E{8}$ & $5.0\E{8}$ & $3.0\E{15}$ & yes \\ % & [1]\\
(b) & Dark clouds         & $1.5\E{19}$ & $10$ & $5.0\E{8}$ & $5.0\E{8}$ & $1.5\E{15}$ & yes \\ % & [1]\\
(c) & Clumps              & $1.0\E{19}$ & $10$ & $1.0\E{9}$ & $5.0\E{8}$ & $1.1\E{15}$ & yes \\ % & [1]\\
(d) & Bok globules        & $1.2\E{18}$ & $10$ & $4.0\E{10}$ & $5.0\E{8}$ & $1.7\E{14}$ & yes \\ % & [1]\\
(e) & Dense cores         & $1.5\E{17}$ & $10$ & $1.0\E{10}$ & $5.0\E{8}$ & $3.4\E{14}$ & yes \\ % & [1]\\
(f) & Hot cores           & $1.0\E{17}$ & $200$ & $1.0\E{14}$ & $5.0\E{8}$ & $1.5\E{13}$ & yes \\ % & [1]\\
\end{tabular}\\
$R$ -- cloud radius; $T$ -- cloud temperature; $n_0$ -- cloud density; $v$ -- cloud speed; $r_0$ -- derived cloud scale height; Ablate? -- If cloud and jet parameters (Tab.~\ref{tab:jetparam}) fulfill Eqs.~(\ref{eq:requiredbulk}) and (\ref{eq:shockspeed1}) 
%References: [1] \cite{co07}, [2] \cite{dea99}, [3] \cite{p06} {\bf MAYBE REMOVE???}
\label{tab:realclouds}
\end{table*}
\begin{figure*}
\begin{minipage}{0.32\linewidth}
\centering \resizebox{\hsize}{!}
{\includegraphics{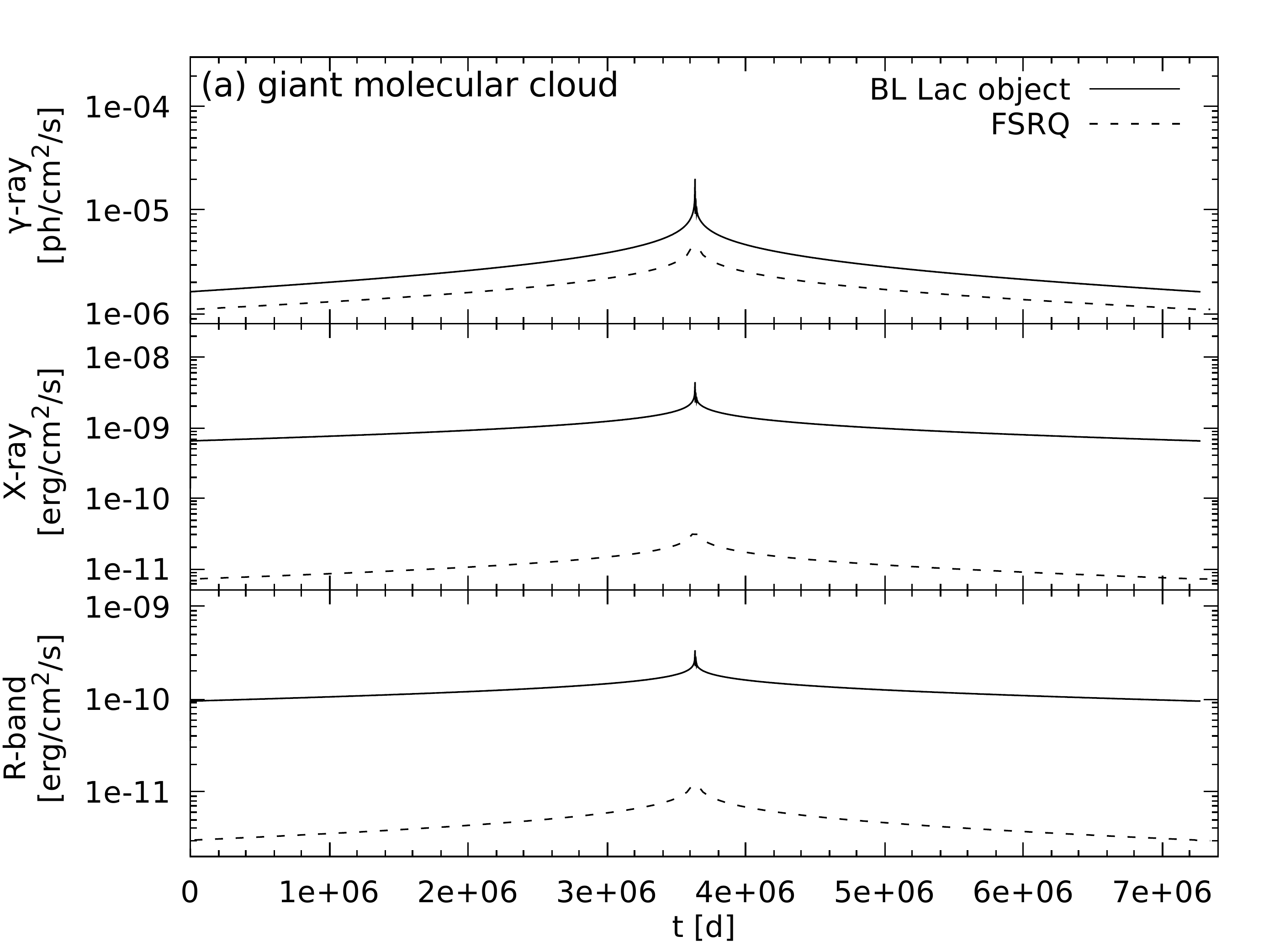}}
\end{minipage}
\hspace{\fill}
\begin{minipage}{0.32\linewidth}
\centering \resizebox{\hsize}{!}
{\includegraphics{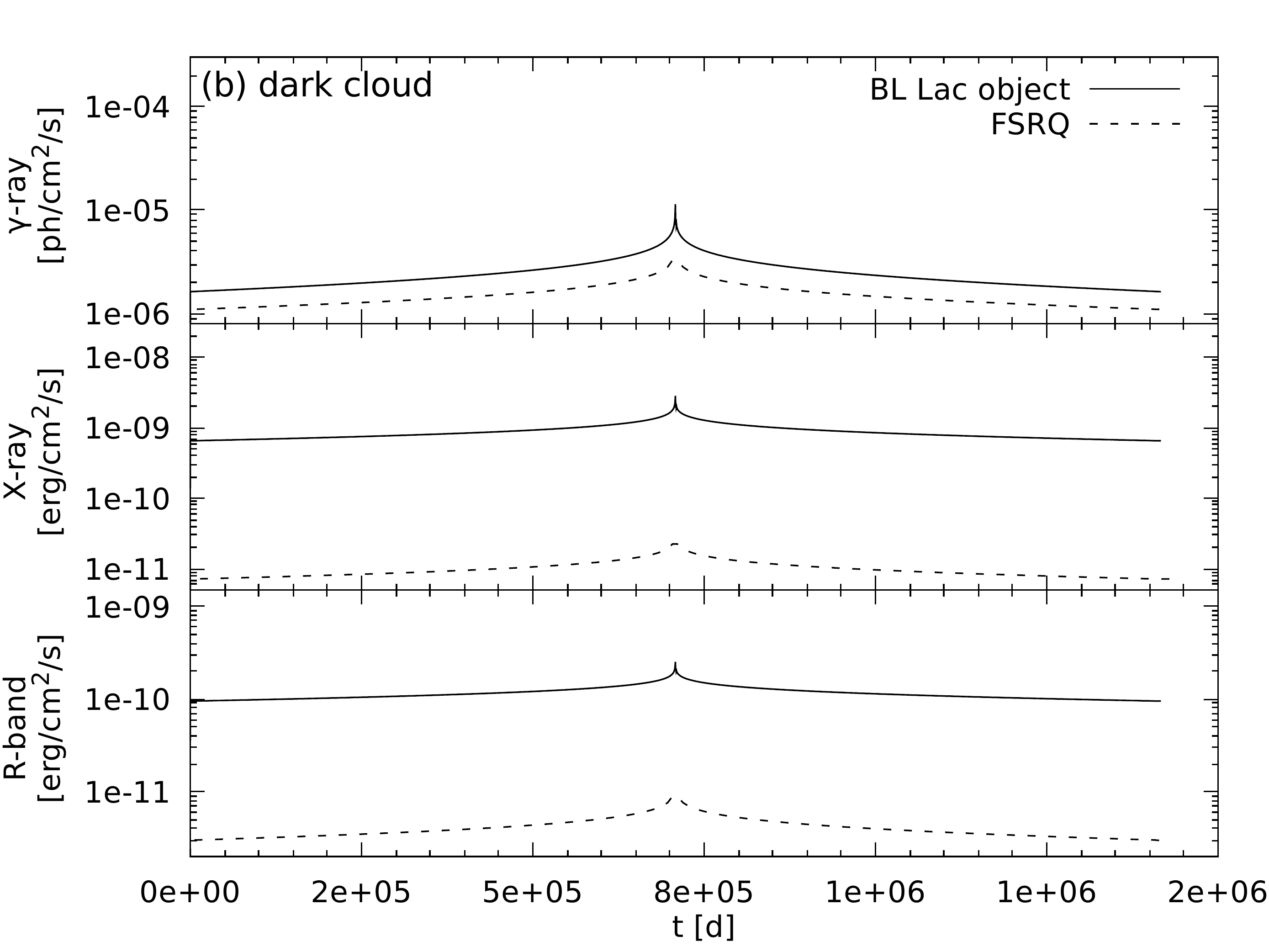}}
\end{minipage}
\hspace{\fill}
\begin{minipage}{0.32\linewidth}
\centering \resizebox{\hsize}{!}
{\includegraphics{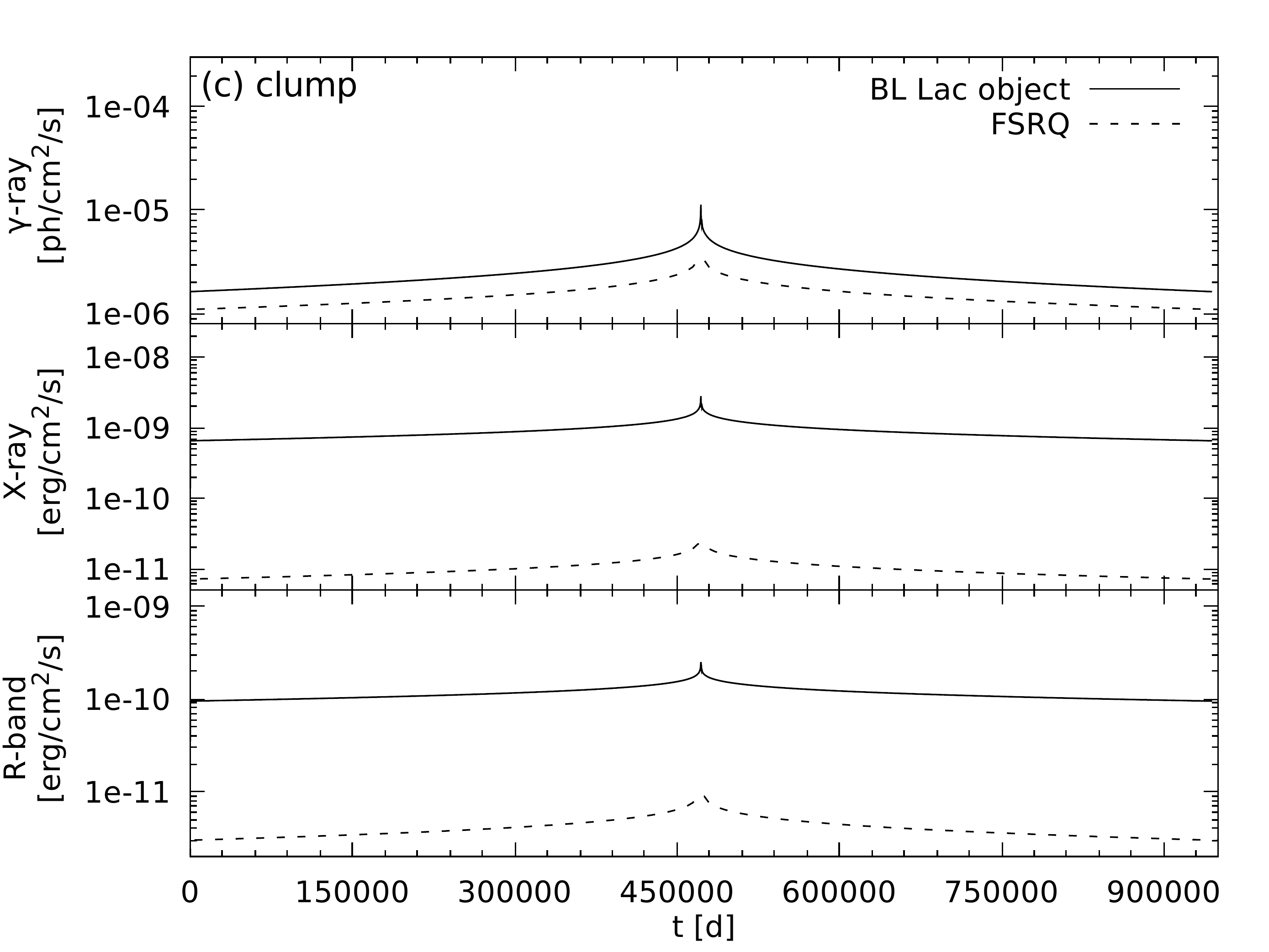}}
\end{minipage}
\newline
\begin{minipage}{0.32\linewidth}
\centering \resizebox{\hsize}{!}
{\includegraphics{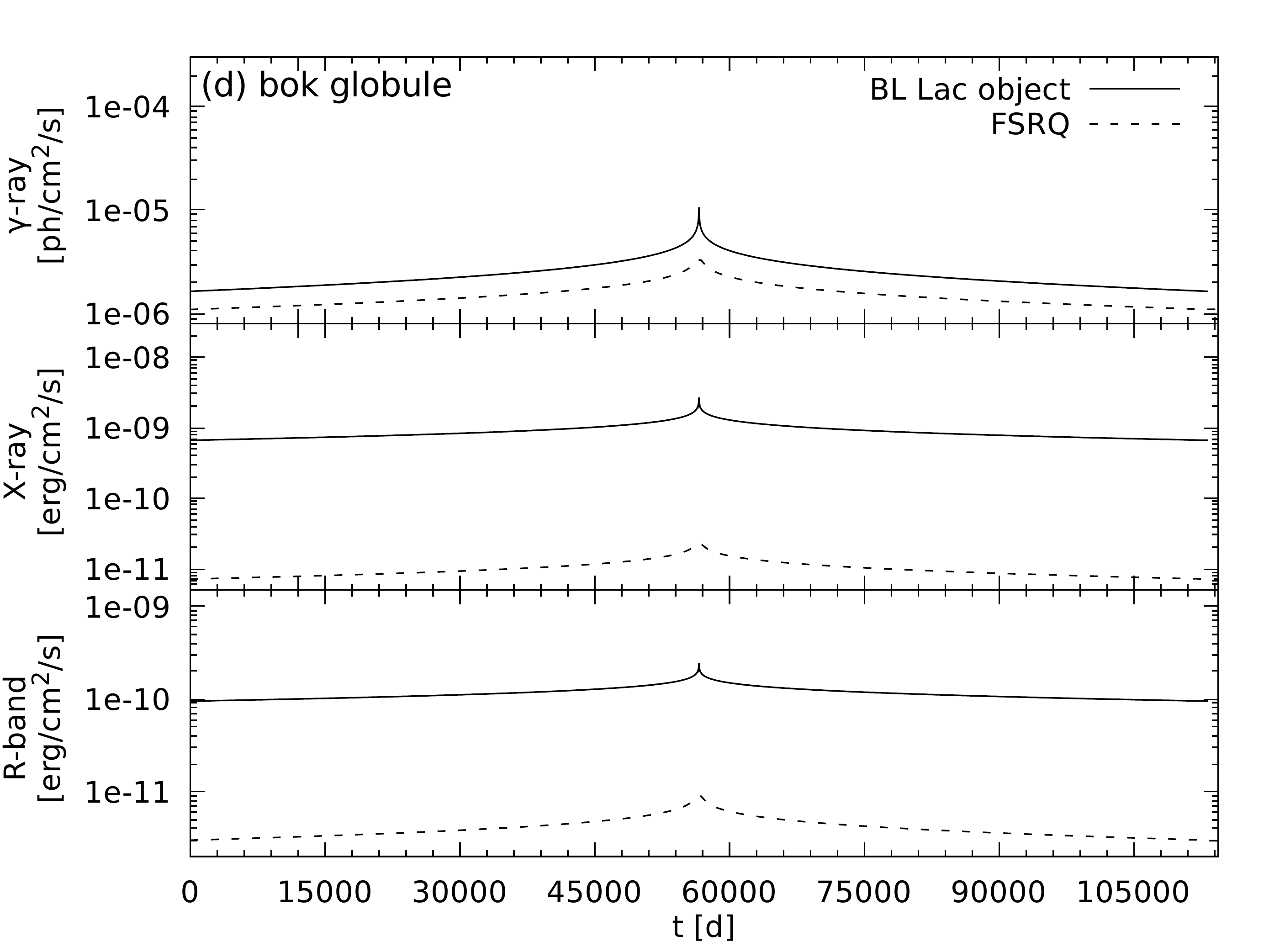}}
\end{minipage}
\hspace{\fill}
\begin{minipage}{0.32\linewidth}
\centering \resizebox{\hsize}{!}
{\includegraphics{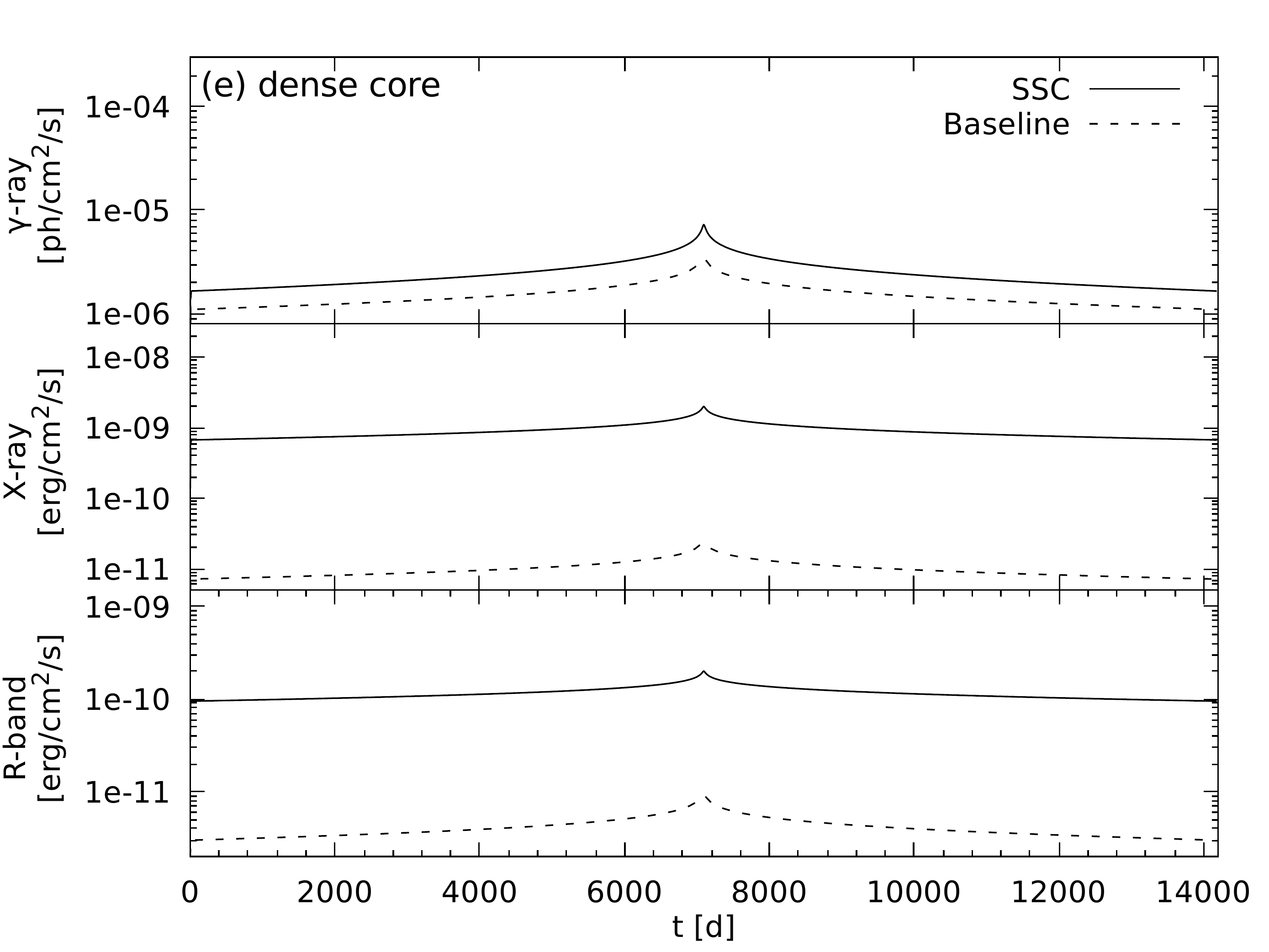}}
\end{minipage}
\hspace{\fill}
\begin{minipage}{0.32\linewidth}
\centering \resizebox{\hsize}{!}
{\includegraphics{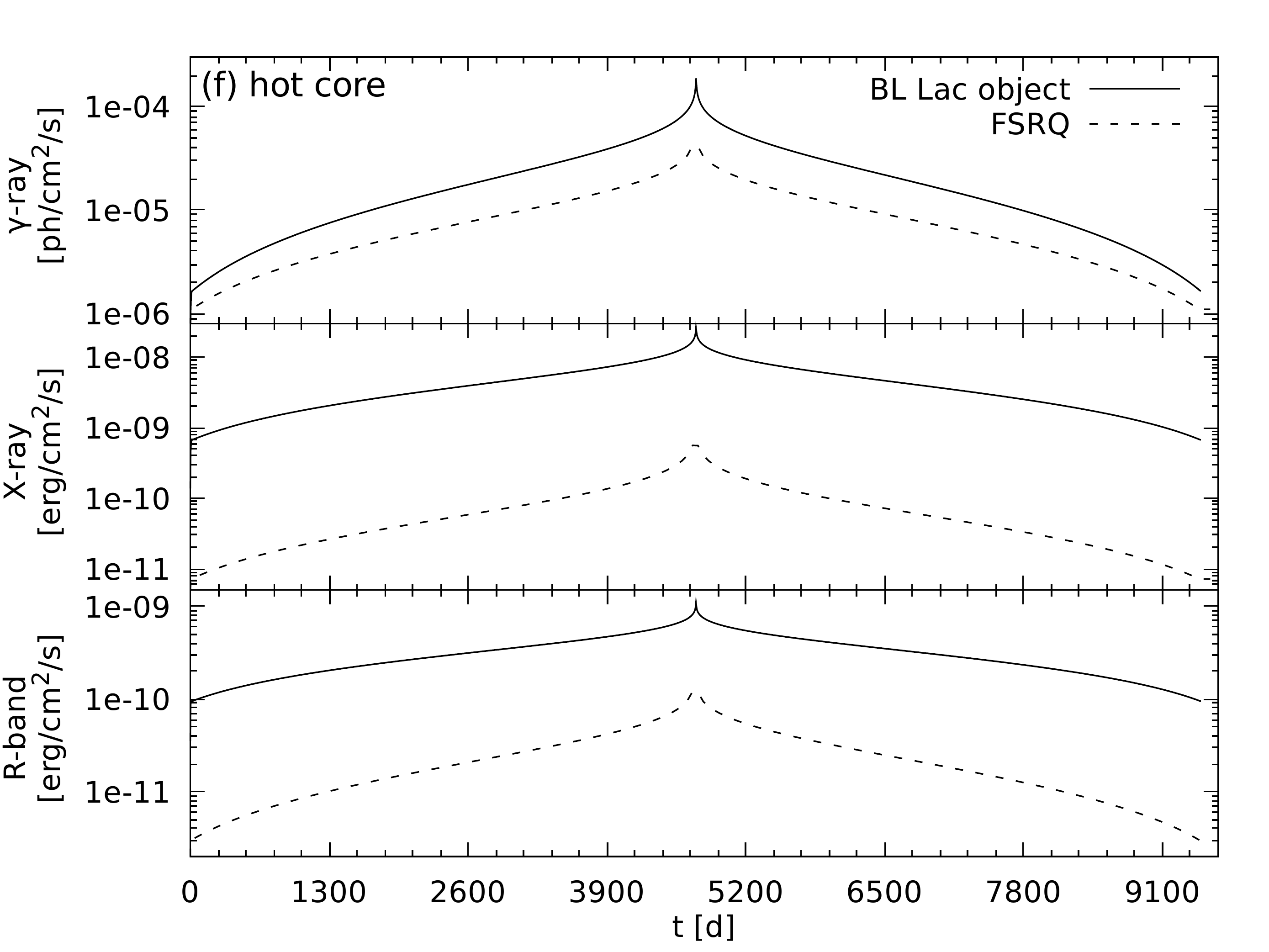}}
\end{minipage}
\caption{Lightcurves of the FSRQ (dashed) and BL Lac object (solid) cases in the observer's frame for three energy bands for examples of real clouds: (a) Giant molecular cloud, (b) dark cloud, (c) clump, (d) bok globule, (e) dense core, and (f) hot core. The jet parameters are given in Tab.~\ref{tab:jetparam}, and the cloud parameters are provided in Tab.~\ref{tab:realclouds}. Note the logarithmic y-axes.
}
\label{fig:realclouds}
\end{figure*} 
Before we proceed, a note is required on the free parameters. As we have used the scale height as a free parameter in the previous section, we implicitly took the cloud temperature $T$ as a dependent variable. However unlike the scale height, $T$ is a measurable quantity, and therefore the scale height becomes the dependent variable from now on. Interestingly, the injection rate, Eq.~(\ref{eq:pratecomov}), is proportional to $n_0r_0^2$. Inserting Eq.~(\ref{eq:scaleheight}), this becomes $n_0r_0^2=\tilde{c}T$, independent of the density. Hence, the influence of the density on the lightcurves is minor. The injection is therefore driven by the speed and the temperature of the cloud. 

With this in mind, we can discuss the lightcurves from interstellar clouds. Some cloud types and their typical parameters \citep{co07} are given in Tab.~\ref{tab:realclouds}. Clearly, this list is not exhaustive, and should be considered as examples. All these clouds fulfill the ablation conditions in Eqs.~(\ref{eq:requiredbulk}) and (\ref{eq:shockspeed1}). 

As we have seen in the previous section, the speed of the cloud has an enormous influence on the resulting lightcurve. Hence, stronger flares may be expected for clouds relatively close to the black hole. Given that many AGN are located in elliptical hosts, the nuclear activity may be a result of the merger of galaxies and gas and dust (in the form of clouds) is pushed into the galactic center. In turn, many clouds will come close to the black hole and the jet while moving rapidly. This provides the necessary ingredients for our model. We assume that the clouds have reached a distance to the black hole within the radius of the BLR, and hence move with roughly the orbital speed of the BLR -- about $5000\,$km\,s$^{-1}$. This allows us to use the FSRQ model. Additionally, as the BL Lac object scenario amplifies the variability, we also provide the corresponding lightcurves.

The resulting lightcurves are shown in Fig.~\ref{fig:realclouds} with dashed lines for the FSRQ parameters and solid lines for the BL Lac object model. Obviously, all cases exhibit strong flares, varying in duration and peak flux. The evolution of the lightcurve also changes slightly in accordance with the discussion of the previous section. The duration is, of course, governed by the size of the objects (and they are ordered in decreasing size), so the duration drops from case to case. The peak flux in turn depends on the particle number, which depends on both the size and the density. For the FSRQ parameter set, the flux variation is on the order of a few in all three bands. In the BL Lac object case, the statements of the previous section hold that the variation in the \g-ray band is roughly quadratically the variation in the X-ray and R bands. That is, if variations in the X-ray and R bands are on the order of 1 order of magnitude, the \g-ray band exhibits variations on the order of 2 orders of magnitude. The most notable variation takes place in ''hot cores``, which is the densest and hottest of the examples. In this case, the lightcurve varies more than an order of magnitude even in the FSRQ case.

Astrospheres of RGB stars are another common ``cloud'' type in elliptical galaxies. We discuss their case in the appendix as they are not fully compatible with our assumptions on the derivation of the cloud's density structure.

%
%#######################################################################################################################################
%#######################################################################################################################################
%
\section{Discussion} \label{sec:sumcon}
Elliptical galaxies -- the hosts of blazars -- form through the collision of gas-rich spiral galaxies. Much of the free gas is funneled into the center of the galaxy, where an AGN is turned on. Clouds of gas will enter the galactic center out of any direction, and may encounter the relativistic jet. We have considered such an encounter of a cloud with the jet as a particle injection process that will produce long-lasting flares.

We first derived the analytical equations that describe the injection function of a spherical, isothermal gas cloud that is only shaped by the hydrostatic equilibrium between its self-gravity and its gas pressure. This expanded the work of \cite{zea17} and provides the correct normalization of the injection process. We also derived the necessary jet conditions to fully ablate the cloud. It turns out that jets should be able to ablate most cloud types. We then proceeded to study the theoretical lightcurve shapes considering various parameter sets for both the cloud and the jet. The most important cloud parameter in this regard is the scale height, which depends on the temperature and the central density of the cloud. The scale height's value relative to the cloud radius determines the homogeneity of the cloud. Homogeneous clouds, i.e. those with a large scale height, produce round lightcurves with a steep rise/decay and relatively flat maximum, while clouds with a small scale height produce a peaked lightcurve with a flat rise/decay and a pronounced central peak. As the density of the cloud can be determined from the peak flux of the lightcurve, the shape of the lightcurve gives a strong indication of the temperature of the cloud.

While the jet parameters can be deduced from observations before the flare, they also have a significant influence on the lightcurve. Most notably, parameters describing BL Lac objects produce a significantly larger variability than FSRQ parameters. While this changes the lightcurve shape, it has no major influence on the peak-structure, and therefore on the possibility to determine the scale height. This is important, as it does not influence the predictive power of the model. 

Subsequently, we used examples of different cloud types that may be present in an active galaxy, and which may penetrate the jet. Each example results in variable fluxes with different magnitudes, different lightcurve shapes, and variations on different time scales. For the example clouds, the variations take thousands to millions of days, which is obviously too long for proper observations. However, there are much smaller clouds present in the Universe, such as the very dense structures around forming stars. Furthermore, the peaks in Fig.~\ref{fig:realclouds} are quite pronounced in terms of flux variation and duration, lasting only for 1 or 2\% of the entire high state duration. This increases the observational potential. 

One of our major assumptions is that all clouds abide the same hydrostatic density structure. While this is a relatively simple structure, real clouds are much more complex. Especially in star forming regions the cloud structure will be chaotic \citep{kfb20,xl20}, and shaped from gravitational encounters, stellar winds, magnetic fields, etc. It is quite likely that such non-spherically-symmetric structures produce flares that are asymmetric. Furthermore, one can also expect a more complicated density structure with several cores, and turbulent behavior. Additionally, we have treated each cloud as an individual entity. In fact, many of the considered cloud types are part of star-forming regions, and are intertwined. Such a multi-cloud model might produce bright flares on top of an extended high state. These are intriguing possibilities for further applications of the model. In any case, the lightcurve would become more complicated with several peaks. A similar result would be obtained, if several individual clouds would interact with the jet at the same time \citep[e.g.,][]{pbr19}. While this increases the number of ablated particles and, thus, the flux variation, one would observe again several peaks. Disentangling the different clouds may be a complicated endeavor.

In our simulations we have assumed that every particle of the cloud enters the jet. This is unlikely to happen. The interaction of the -- loosely bound -- cloud with the highly energetic jet will result in the ejection of cloud material, and only a fraction of the cloud will enter the jet. It is difficult to quantify how much material is lost in this way, and would require dedicated (M)HD simulations, which are beyond the scope of this paper. Additionally, in case of real clouds, it is unlikely that all particles enter the jet independently of the cloud size with respect to the jet size. If the cloud is much larger than the jet, parts of the cloud will skip the jet. However, if the central and densest part of the cloud enters the jet, the effect should be minor. In fact, in all our examples the scale height is smaller than the jet radius. Despite these issues, a significant amount of particles enters the jet to produce an equally significant flux variation.

As radiation processes we have considered leptonic synchrotron and inverse-Compton emission, as this is the standard blazar emission scenario. However, as the cloud naturally contains protons and heavier nuclei, hadronic radiation processes might be an interesting alternative to produce the emission. On the one hand, the cloud's nuclei may be accelerated to non-thermal speeds and produce radiation on their own \citep[proton synchrotron, pion and muon synchrotron, etc; e.g.][]{bea13,zea19}. On the other hand, the cloud's nuclei could also serve as targets for the jet's relativistic protons, resulting in proton-nucleus interactions and subsequent radiation production \citep{hea20}. This is an intriguing possibility to produce a flare without the need to accelerate the cloud particles. Here, as well, dedicated simulations may provide further insights.

In summary, we have demonstrated that the cloud ablation process is a viable option to produce long-lasting high states in blazars. Further studies, especially (M)HD or particle-in-cell simulations of the entire process, are strongly encouraged.

%
%#######################################################################################################################################################################
%_______________________________________________________________________________________________________________________________________________________________________
%
\section*{Acknowledgement}
The authors wish to thank Markus B\"ottcher, Patrick Kilian, Klaus Scherer, Valent\'i Bosch-Ramon, Maxim Barkov, Frank Rieger, and Kerstin Wei\ss{} for stimulating discussions on model and manuscript details, as well as cloud parameters. 
We also thank the anonymous referee for valuable suggestions that helped to improve the manuscript. 
Funding by the German Ministry for Education and Research (BMBF) through grant 05A17PC3 is gratefully acknowledged.
%
%#######################################################################################################################################################################
%_______________________________________________________________________________________________________________________________________________________________________
%

%
%#######################################################################################################################################################################
%_______________________________________________________________________________________________________________________________________________________________________
%
\begin{appendix}
\section{The astrosphere of RGB stars} \label{app:RGB}
\begin{figure}
\centering
\includegraphics[width=0.48\textwidth]{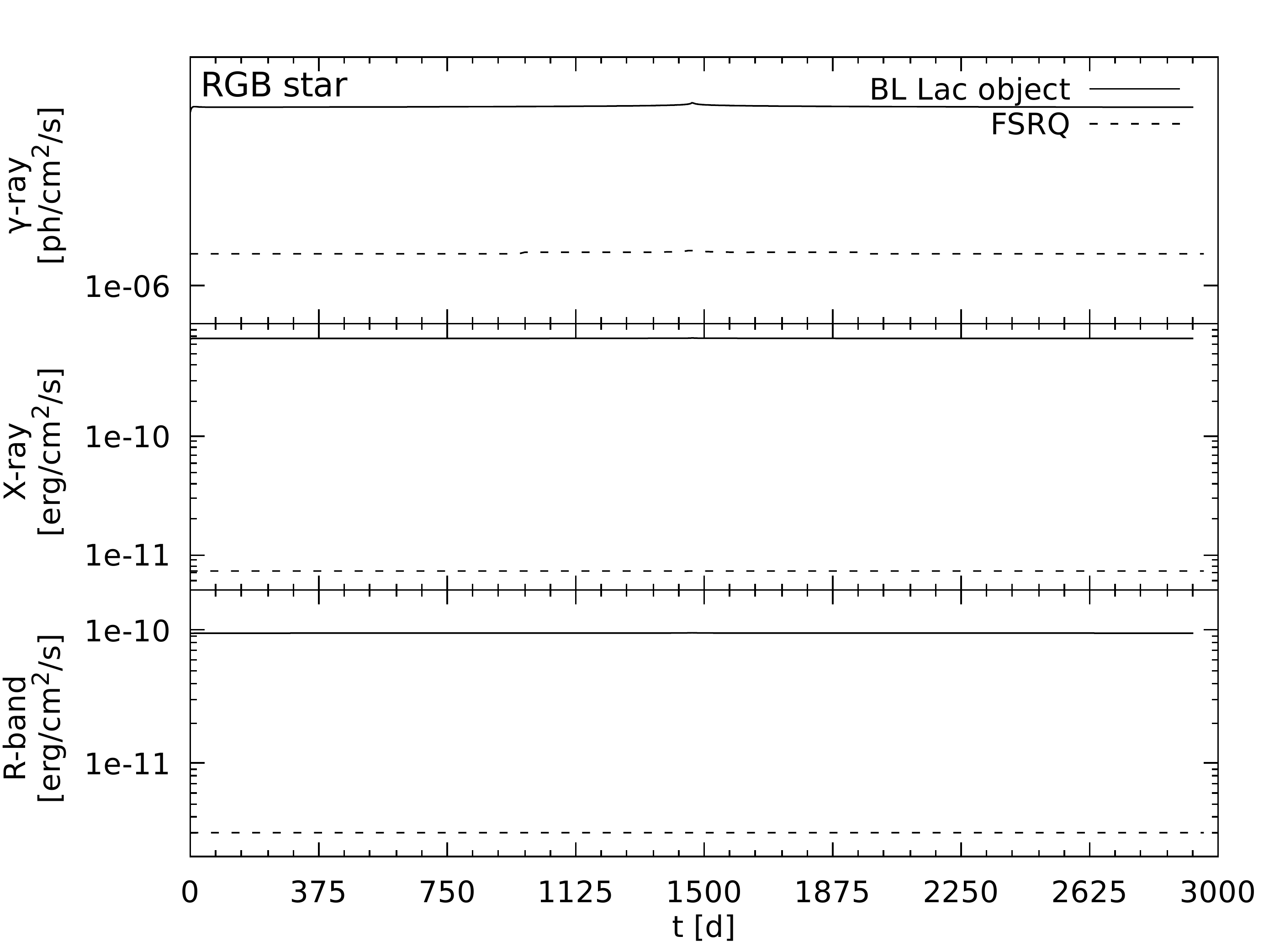}
\caption{Lightcurves for the FSRQ (dashed) and BL Lac object (solid) cases in the observer's frame for three energy bands for the astrosphere of an RGB star. See text for parameters. Note the logarithmic y-axes.
}
\label{fig:RGB}
\end{figure}
The astrosphere of RGB stars is a special case in our study, as they are blown up by the stellar wind, i.e. an outward stream of particles. In turn, our assumption of an isothermal cloud is a poor representation of such an object. Additionally, the gravitational pull keeping the astrosphere together is provided by the star and not the self-gravity of the cloud. This means that the scale height does not follow from the considerations in Sec.~\ref{sec:ablation}, and in fact loses its meaning. However, we can use the known density structure of astrospheres, which is similar to the hydrostatic case. From some distance $r_{s,0}$ above the star's surface up to the termination shock the density follows an inverse-square law \citep[e.g.,][]{sea20}. Following our deliberations in Sec.~\ref{sec:ablation}, we assume that the star remains intact during the interaction with the jet. Hence, we can replace the scale height with $r_{s,0}$ and our formalism can be applied.

Known parameters of RGB star winds are the mass loss rate $\dot{M}$ \citep{sc05,oea07}, the wind velocity $v_w$ \citep[e.g.,][]{rcb98}, and the radius of the star $R_s$ \citep[e.g.,][]{vl09}. With these parameters we can calculate the central density as

\begin{align}
    n_0 &= \frac{\dot{M}}{4\pi R_s^2 v_w m_p} \nonumber \\
    &= 8.2\E{10}\,\mbox{cm}^{-3}\, \est{\dot{M}}{10^{-6}\,M_{\odot}\,\mbox{yr}^{-1}}{} \est{R_s}{50\,R_{\odot}}{-2} \est{v_w}{30\,\mbox{km}\,\mbox{s}^{-1}}{-1} \label{eq:RGBn0}.
\end{align} 
The termination shock distance depends on the density $n_{\rm ISM}$ and speed $v_{\rm ISM}$ of the interstellar medium of the distant galaxy, for which reasonable assumptions can be made. Note that $v_{\rm ISM}$ is a relative speed between the interstellar medium and the star, while the speed of the star (cloud) used in our simulation is the speed of the star (cloud) penetrating the jet. From hydrodynamical considerations, the termination shock distance $R_{s,t}$ can be calculated as \citep{p58,w00}

\begin{align}
    R_{s,t} &= \sqrt{\frac{\dot{M}v_w}{4\pi m_p n_{\rm ISM}v_{\rm ISM}^2}} \nonumber \\
    &= 3.1\E{16}\,\mbox{cm}\, \est{\dot{M}}{10^{-6}\,M_{\odot}\,\mbox{yr}^{-1}}{1/2} \est{v_w}{30\,\mbox{km}\,\mbox{s}^{-1}}{1/2} \nonumber \\
    &\quad\times \est{n_{\rm ISM}}{100\,\mbox{cm}^{-3}}{-1/2} \est{v_{\rm ISM}}{100\,\mbox{km}\,\mbox{s}^{-1}}{-1} \label{eq:Rtermshock}
\end{align}
which corresponds to about 2000\,AU. Using $r_{s,0} = 0.5\,$AU provides the result shown in Fig.~\ref{fig:RGB}.

The mass loss rate $\dot{M}$ used in Eqs.~(\ref{eq:RGBn0}) and (\ref{eq:Rtermshock}) is at the upper end of RGB mass loss rates, which may be achieved only during a short period of time close to the end of the RGB phase of the star. Note that a reduction in $\dot{M}$ reduces both the central density and the termination shock distance.

Therefore, the result shown in Fig.~\ref{fig:RGB} provides an upper limit case on what could be achieved in this scenario. Apparently, the astrosphere does not contain enough particles to provide a meaningful flare. While a very minor variation is visible in the center of the lightcurve, this tiny variation is not observable within the usual fluctuations of blazar lightcurves.The presence of numerous stars in the jet at the same time \citep{abr13,vtb17} may provide sufficient particles to achieve a more pronounced flare. 

Wolf-Rayet stars exhibit even greater mass loss rates than RGB stars with $\dot{M}_{WR}\sim 10^{-4}$. They might provide sufficiently dense winds in order to produce meaningful flares. However, their total number in a galaxy are very low, and the odds of one interacting with the jet are even lower.

\end{appendix}
\end{document}